\documentclass[10pt, 3p]{elsarticle}

\usepackage[british]{babel}
\usepackage[utf8]{inputenc}
\usepackage{graphics}
\usepackage{graphicx}
\graphicspath{{.},{figures/}}
\usepackage{amsmath,amssymb,esint}
\usepackage{lineno}
\usepackage{multirow}
\usepackage{subfig}
\usepackage{natbib}
\usepackage{eucal}
\usepackage{mathtools}
\usepackage{placeins}
\usepackage{enumerate}
\usepackage[usenames,dvipsnames]{xcolor}
\usepackage[colorlinks]{hyperref}
\usepackage{setspace}
\usepackage{MnSymbol}
\usepackage[linesnumbered,ruled,vlined]{algorithm2e}
\usepackage{array}
\usepackage{cancel}
\usepackage{booktabs}

\biboptions{sort&compress}
\makeatletter
\def\ps@pprintTitle{%
  \let\@oddhead\@empty
  \let\@evenhead\@empty
  \let\@oddfoot\@empty
  \let\@evenfoot\@oddfoot}
\makeatother

\begin{document}

\begin{frontmatter}

\title{An integral surface tension scheme for three--dimensional front tracking frameworks}

\author[affil1]{Gabriele Gennari}
\author[affil1]{Berend van Wachem\corref{cor1}}
\ead{berend.vanwachem@ovgu.de}

\address[affil1]{Chair of Mechanical Process Engineering,
Otto-von-Guericke-Universit\"{a}t Magdeburg,\\ Universit\"atsplatz 2, 39106 Magdeburg, Germany}

\cortext[cor1]{Corresponding author: }

\begin{abstract}
Surface tension plays a central role in many two--phase flow configurations and accurate numerical schemes are crucial to predict the effect of this interfacial force on the fluid system. The integral formulation of the surface tension term, originally proposed by \citet{Popinet1999}, emerges as a natural way to discretise such a term and results in a scheme that conserves momentum both locally and globally, and requires no modifications for two--phase systems with variable surface tension coefficients, such as Marangoni flows. However, to the best of the authors' knowledge, only two--dimensional integral formulations have been proposed in the literature to date. The main reason for the lack of three--dimensional schemes lies in the difficulty of achieving a robust three--dimensional implementation, given the extremely complex shapes that can characterise an interface in a generic 3D flow. In this work, we present the first three--dimensional integral surface tension scheme and we implement it in a sharp front tracking framework. The numerical method is tested across a variety of two--phase flow configurations, such as spherical droplets in static/translating equilibrium, oscillating droplets, thermocapillary motion, and rising bubbles. The integral method is compared against analytical solutions and experimental data, and is benchmarked against well--known schemes, such as the continuous surface force (CSF) and smoothing--based methods for surface tension. The proposed scheme results in comparable spurious velocities with the CSF scheme, but achieves superior accuracy across all the other test cases, with the largest gains observed for droplets oscillating at low Ohnesorge numbers, flows with variable surface tension and rising bubbles undergoing large deformations. The proposed scheme reduces errors in the terminal velocity of a thermocapillary--driven droplet by up to an order of five compared to smoothing--based methods, and overall predicts steady--state shapes of rising bubbles significantly closer to experimental observations, especially for low Morton numbers.
\end{abstract}
\end{frontmatter}

\section{Introduction}
\label{sec:Introduction}
\begin{figure}[!htbp]
    \centering
    \includegraphics[]{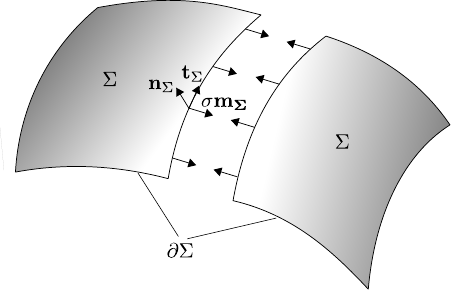}
    \caption{The surface tension force in Young's membrane model.}
    \label{fig:surface_tension}
\end{figure}

Two--phase flows are ubiquitous and occur in a variety of applications, ranging from environmental systems to industrial processes. In many cases, a two--phase system consists of a disperse phase (e.g., droplets, or bubbles) surrounded by a continuous medium. This is the case, for example, of bubble column reactors \citep{Kluytmans2001}, flotation cells \citep{Tiedemann2025}, and air entrainment/gas transfer at the ocean free--surface \citep{Deike2016}. In these flow configurations, a central role is played by the surface tension effect, which originates from intramolecular forces \citep{Berg2009}. One of the most evident consequences of surface tension is the fact that an undisturbed droplet (e.g., oil droplet immersed in a liquid with same density) assumes a spherical shape, as this configuration minimises the surface energy for a given volume of liquid. From the point of view of continuum mechanics, the interface between two phases is represented as a membrane, i.e., an infinitely thin, mass--less, surface that separates the two fluids. Under this model, known as Young's membrane model after the work of \citet{Young1805}, surface tension is a force per unit length that pulls the interface $\Sigma$ along its perimeter $\partial \Sigma$ (see Figure \ref{fig:surface_tension}).
It acts along the direction of the unit vector $\mathbf{m}_\Sigma$, which corresponds to the cross product between the normal at the interface $\mathbf{n}_\Sigma$ and the tangent vector along its perimeter $\mathbf{t}_\Sigma$, i.e., $\mathbf{m}_\Sigma = \mathbf{t}_\Sigma \times \mathbf{n}_\Sigma$. Surface tension is expressed as $\sigma \mathbf{m}_\Sigma$, where the surface tension coefficient $(\sigma)$ represents the magnitude of such an interfacial force. A complete different behaviour from the static droplet in a spherical equilibrium can be observed when $\sigma$ is not uniform along the interface. This typically happens when the system is non--isothermal, as $\sigma$ decreases with temperature, or in the presence of contaminants, such as impurities or surface--active substances. In this case, the two--phase system cannot reach a static equilibrium state and a flow field develops. This effect is known as Marangoni flow, after the pioneering work of \citet{Marangoni1871}. Examples of Marangoni effects include active droplets \citep{Chen2024a, Kant2024} and evaporating sessile droplets \citep{Jain2025a, Jain2024}.  

Given the vast amount of applications of two--phase flows with surface tension effects, remarkable effort has been put in the development of accurate numerical schemes to predict surface tension effects. For interface capturing approaches, such as volume of fluid (VOF) or level set (LS), the de facto standard of surface tension modelling is the continuous surface force (CSF) scheme presented in the work of \citet{Brackbill1992}. This model is based on a volumetric formulation of the corresponding source term and requires a numerical discretisation of the interfacial $\delta$--function, which makes it inherently mesh--dependent and impossible to achieve a sharp representation of the force. Additional challenges arise for two--phase systems where Marangoni flows are present, for which the surface gradient of $\sigma$ (i.e., $\nabla_\Sigma \sigma$) appears in the source term. As shown in Figure \ref{eq:surf_tension}, surface tension is an internal force and, by definition, the integral over the whole interface must be null. However, the CSF scheme does not satisfy this constraint, neither locally nor globally \citep{Popinet2018}. The main reason behind its popularity lies in the fortunate possibility of reformulating the source term as the gradient of a potential (the volume fraction field), which leads to a well--balanced implementation and allows to recover equilibrium solutions when pressure gradients and surface tension balance out. This aspect has been proved in several works that show almost null spurious velocities after approximately one viscous time unit around a static droplet in equilibrium when symmetric boundary conditions are applied (i.e., only a quarter or half of the sphere is modelled) \citep{Popinet2009, Abadie2015}. Such accurate results require, however, a very accurate estimation of the curvature, for which height functions is a typical approach \citep{Popinet2018, Popinet2009, Evrard2020}.

In front tracking (FT) approaches \citep{Tryggvason2001}, the interface is treated explicitly in a Lagrangian fashion and typically consists of a triangular surface mesh. This framework allows for a direct computation of the surface tension effect at the interface, and overcomes some of the limitations of the CSF scheme, such as the discretisation of the $\delta$--function, and adapts seamlessly to non--uniform surface tension coefficients. However, the classic approach typically adopted in FT approaches is based on the spreading of the surface tension term from the interface onto the Eulerian mesh (i.e., the grid where the governing equations are solved), using a smoothing kernel \citep{Tryggvason2001, Gorges2022, Shin2002, Dijkhuizen2010, Deen2004} and, therefore, does not result in a truly sharp description of the interfacial force; this approach will be referred to as classic FT in the following. Hybrid formulations that combine the CSF method with information computed at the Lagrangian level have also been proposed \citep{Shin2005, Shin2011, Gorges2025b}, but suffer from the same additional complexity as CSF when non--uniform surface tension coefficients are present. Obviously, also standard CSF methods can be combined with FT frameworks \citep{Gorges2025b}.
 
Owing to the explicit representation of the interface, the FT approach offers a natural framework to compute the surface tension force at the Lagrangian level. However, in order to achieve a consistent representation of the interfacial force on the Eulerian mesh, a sharp distribution of such as source term must be implemented. \citet{Popinet1999} propose an integral method for a 2D FT framework, which consists of computing the intersections between the front and the borders of a computational cell, and applying the resulting surface tension force (acting along the perimeter of the intersection) to the RHS of the momentum equation for the same cell. A pressure correction scheme is implemented when computing the gradient of pressure for cells cut by the interface (mixed cell). In this case, the pressure field is discontinuous and a better estimation of the average mixed face pressure is obtained from a weighted average of the pressure values from both sides of the interface. 

A conceptually similar approach is proposed for the volume of fluid method \citep{Baltussen2014, Thuy-Petrov2026}, where the surface tension force is computed in a Lagrangian way from the reconstructed VOF interface and distributed onto the computational grid. Since the VOF--reconstructed interface does not guarantee continuity of the interface across adjacent cells, extra care must be taken into account to mitigate spurious effects resulting from piecewise discontinuous interfaces. A pressure correction method is also implemented, but with the aim of reducing the magnitude of the surface tension force by subtracting the average pressure jump across the interface. As a result, the computed pressure field does not reflect the actual physical solution.

The integral scheme of \citet{Popinet1999} is further developed in the works of \citet{Abu-Al-Saud2018} and \citet{Saini2025a}. In the first work, the authors implement the integral formulation in a 2D level set framework, and introduce a pressure correction scheme that explicitly takes into account the Laplacian pressure jump across the interface. The work of \citet{Saini2025a} builds on the previous one and extends the method to a 2D VOF approach. The integral formulation offers several advantages, since it guarantees that the surface tension source term is locally and globally conservative, and does not require a numerical discretisation of the interfacial $\delta$--function. Furthermore, this scheme does not require any modifications for problems with variable surface tension coefficients. However, it is not possible to formulate this term as the gradient of a potential function, making a well--balanced implementation not always achievable. The authors in \citep{Popinet1999, Abu-Al-Saud2018, Saini2025a} show very good performance of the integral formulations for 2D test cases, but conclude that an extension to 3D problems is not trivial due to the complexity in computing intersections between three--dimensional interfaces and the computational cells. To the best of the authors' knowledge, no previous attempts at implementing the integral scheme in a 3D framework have been made. In this work we present the first implementation of the integral scheme in our recently proposed sharp 3D FT framework \citep{Gorges2025b},which relies on a piecewise-parabolic interface calculation (PPIC) approach to obtain a sharp representation of the volume fraction field. Smoothing--based techniques typically employed in FT to spread interfacial properties onto the Eulerian mesh are therefore avoided. The interested reader is referred to the work of \citet{Gorges2025b} for more details.

The rest of the work is organised as follows. A detailed description of the integral surface tension scheme and its implementation into a FT framework is discussed in section \ref{sec:The surface tension force}, where also classic FT and CSF are briefly introduced, as they will serve as benchmark solutions. The Navier--Stokes and FT solvers are introduced in section \ref{sec:Governing equations and flow solver}, whilst a detailed validation of the proposed method is presented in section \ref{sec:Validation} across several test cases, which include both analytical and experimental solutions. Finally, the conclusions are drawn in section \ref{sec:Conclusions}.

\section{The surface tension force}
\label{sec:The surface tension force}
As discussed in section \ref{sec:Introduction}, two main approaches for the treatment of surface tension in two--phase flows simulations are available, namely the integral and volumetric formulations, which are formally described in section \ref{sec:ntegral and volumetric formulations}. The novel integral surface tension scheme proposed in this work is discussed in detail in section \ref{sec:The integral surface tension scheme and its implementation into a front tracking framework}. We will compare this scheme against two benchmarks, the classic FT and the CSF methods, which are briefly introduced in sections \ref{sec:The classic approach in front tracking: a smoothing--based formulation} and \ref{sec:The continuous surface force scheme}, respectively.  

\subsection{Integral and volumetric formulations}
\label{sec:ntegral and volumetric formulations}
\begin{figure}[!htbp]
    \centering
    \includegraphics[]{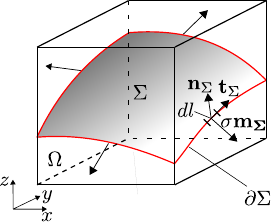}
    \caption{Surface tension force acting on the control volume $\Omega$ along the interface boundary $\partial \Sigma$ (red line).}
    \label{fig:surface_tension_control_volume}
\end{figure}

The global force acting on the control volume $\Omega$ (see Figure \ref{fig:surface_tension_control_volume}) is:
\begin{equation}
    \mathbf{F}_\sigma = \oint_{\partial \Sigma} \sigma \mathbf{m}_\Sigma \,dl\
    \label{eq:surf_tension}
\end{equation}
where $\mathbf{m}_\Sigma$ is the direction of the surface tension force and is computed as $\mathbf{m}_\Sigma = \mathbf{n}_\Sigma \times \mathbf{t}_\Sigma$; the vectors $\mathbf{n}_\Sigma$ and $\mathbf{t}_\Sigma$ are the normal direction and the tangent along the boundary $\partial \Sigma$, respectively.
Equation \ref{eq:surf_tension} is the most straightforward way, from a mathematical point of view, to represent the effect of surface tension. Numerical schemes that implement surface tension in this form are generally referred to as integral formulations \citep{Popinet2018}. This approach inherently conserves momentum locally and globally, similar to flux--based formulations in finite volumes methods, where fluxes across a face shared by two neighbouring cells cancel out exactly. Further advantages include the computation of flows with variable surface tension (e.g., Marangoni flows), since $\sigma$ appears simply as a coefficient in Eq. \ref{eq:surf_tension}. Despite its simple derivation and momentum--conserving properties, only a few works have attempted to discretise surface tension using Eq. \ref{eq:surf_tension}, due to difficulties in computing the geometric intersection $\partial \Sigma$ between the interface and the boundaries of the control volume, as well as in extending the methodology to three--dimensional interfaces.   

Because of the control volume nature of finite volume methods, where governing equations are discretised to satisfy the conservation of variables (e.g., mass and momentum) across closed volumes, the most popular treatment for surface tension is the volumetric formulation. This approach consists in rewriting the line integral in Eq. \ref{eq:surf_tension} as a volume integral. An intermediate surface integral can be immediately written as:
\begin{equation}
    \oint_{\partial \Sigma} \sigma \mathbf{m}_\Sigma \,dl\ = 
    \iint_\Sigma \frac{\partial (\sigma \mathbf{m}_\Sigma)}{\partial s} \,ds\
    \label{eq:surf_tension_surf}
\end{equation}
and, using the first Frenet formula \citep{Tryggvason2011a}, the integrand on the RHS can be rewritten as:
\begin{equation}
    \frac{\partial (\sigma \mathbf{m}_\Sigma)}{\partial s} = 
    \sigma \frac{\partial \mathbf{m}_\Sigma}{\partial s} + \mathbf{m}_\Sigma \frac{\partial \sigma}{\partial s} = 
    \sigma \kappa \mathbf{n}_\Sigma + \nabla_\Sigma \sigma
    \label{eq:frenet}
\end{equation}
where $\kappa$ is the curvature of the interface and $\nabla_\Sigma$ is the surface gradient operator. Inserting Eq. \ref{eq:frenet} into Eq. \ref{eq:surf_tension_surf} and introducing the surface--delta function $\delta_\Sigma$ leads to:
\begin{equation}
    \iint_\Sigma \left( \sigma \kappa \mathbf{n}_\Sigma + \nabla_\Sigma \sigma \right) \, ds\ = 
    \iiint_\Omega \left( \sigma \kappa \mathbf{n}_\Sigma + \nabla_\Sigma \sigma \right) \delta_\Sigma \,dV\
    \label{eq:surf_tension_vol}
\end{equation}
where the RHS is the so--called volumetric formulation, and $\delta_\Sigma$ is the interfacial $\delta$--function. The integrand on the RHS is a volumetric source term, and it will be indicated in the following with the symbol $\mathbf{f}_\sigma \delta_\Sigma = \left( \sigma \kappa \mathbf{n}_\Sigma + \nabla_\Sigma \sigma \right) \delta_\Sigma$.

It appears immediately evident that this formulation has significantly more complexity compared to the integral one (Eq. \ref{eq:surf_tension}), since it requires numerical approximations of $\delta_\Sigma$ and, in case of variable $\sigma$, of the additional term $\nabla_\Sigma$. Both terms are challenging from a numerical perspective, since $\delta_\Sigma$ is a discontinuous function, whereas $\nabla_\Sigma$ is a projection of the gradient operator onto the interface, which is typically not available explicitly in interface capturing methods, such as VOF and LS. A further disadvantage is that the numerical discretisation of the volumetric formulation guarantees neither local nor global momentum conservation. Despite these disadvantages, this approach remains the first choice in numerical simulations, due to the straightforward extension to 3D interfaces and a relatively easy implementation in a well--balanced form, which ensures balancing between pressure gradient and the surface tension force. One of the most popular implementation of the volumetric formulation is the CSF method \citep{Brackbill1992}, which is briefly described in section \ref{sec:The continuous surface force scheme}.  

\subsection{The integral surface tension scheme and its implementation into the front tracking framework}
\label{sec:The integral surface tension scheme and its implementation into a front tracking framework}
In this section we adapt the integral surface tension scheme \citep{Abu-Al-Saud2018, Saini2025a} to the front tracking framework and describe the details of its numerical implementation. In the following we will describe the discretisation of the pressure gradient and surface tension terms along the $x$--direction for a Cartesian mesh; $y$ and $z$--components can be obtained straightforwardly by permutation of the indices.

The contributions of pressure gradient and surface tension force to the $x$--component of momentum are (Figure \ref{fig:surface_tension_control_volume}):
\begin{equation}
    - \iiint_\Omega \frac{\partial p}{\partial x} \,dV\ + 
    \oint_{\partial\Sigma} \sigma \mathbf{m}_\Sigma \cdot \mathbf{x} \,dl\
    \label{eq:momentum-x}
\end{equation}
where $p$ is the pressure and $\mathbf{x}$ is the unit vector along the $x$--direction. The surface tension contribution can be rewritten as:
\begin{equation}
    \oint_{\partial \Sigma} \sigma \mathbf{m}_\Sigma \cdot \mathbf{x} \,dl\ \approx 
    \sum_\mathrm{faces} \left( \sigma l_\Sigma \mathbf{m}_\Sigma \right)_\mathrm{f} \cdot \mathbf{x} 
    \label{eq:surf_tension_integral}
\end{equation}
where $\left( \sigma l_\Sigma \mathbf{m}_\Sigma \right)_\mathrm{f}$ is the average contribution of surface tension along the interface border (with length $l_\Sigma$) on face $f$. Such a contribution is obviously non--null only for the faces cut by the interface (faces with normals $\mathbf{x}, \mathbf{y}$ in Figure \ref{fig:surface_tension_control_volume}), whereas faces that are not crossed by $\Sigma$ simply have $l_\Sigma = 0$.

\begin{figure}[!htbp]
    \centering
    \includegraphics[]{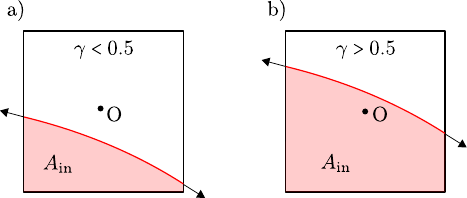}
    \caption{Computation of the surface fraction $\gamma$ for $\gamma < 0.5$ (a) and $\gamma > 0.5$ (b). The red area corresponds to the face region inside the interface $(A_\mathrm{in})$.}
    \label{fig:face_fractions}
\end{figure}

The pressure gradient term in Eq. \ref{eq:momentum-x} can be turned into a surface integral via Gauss' theorem:
\begin{equation}
    \iiint_\Omega \frac{\partial p}{\partial x} \,dV\ = \iint_{\partial \Omega} p \mathbf{x} \cdot \mathbf{n} \,dS\
    \label{eq:gauss_pressure}
\end{equation}
where $\mathbf{n}$ is the outward--pointing normal at the boundary $\partial \Omega$ of the control volume $\Omega$; for a Cartesian mesh, $\partial \Omega$ corresponds to the union of the six faces bounding the cell. As proposed in \citep{Abu-Al-Saud2018, Saini2025a}, an accurate computation of the integral on the RHS of Eq. \ref{eq:gauss_pressure} requires to explicitly take into account the pressure jump across the interface due to surface tension. Depending on the relative intersection between the interface and the faces, two possible scenarios can occur: the face can be mainly outside the interface (Figure \ref{fig:face_fractions}a) or mainly inside (Figure \ref{fig:face_fractions}b). 
To distinguish between these two cases, a face--fraction parameter $\gamma$ is defined as:
\begin{equation}
    \gamma = \frac{A_\mathrm{in}}{A}
    \label{eq:gamma_f}
\end{equation}
where $A$ is the area of the face, whereas $A_\mathrm{in}$ is the area of the face inside the interface. The parameter $\gamma$ is bounded between $0 < \gamma < 1$, and the two configurations immediately follow as $\gamma < 0.5$ (Figure \ref{fig:face_fractions}a) and $\gamma > 0.5$ (Figure \ref{fig:face_fractions}b). In the first case, the integral in Eq. \ref{eq:gauss_pressure} can be discretised as (for a face with normal $\mathbf{x}$):
\begin{equation}
    \begin{aligned}
        \iint_\mathrm{face} p \,dS\ 
        &\approx p_\mathrm{O} (1 - \gamma) A + 
    (p_\mathrm{O} + \kappa \sigma) \gamma A \\
    &= (p_\mathrm{O} + \kappa \sigma \gamma) A
    && \text{for } \gamma \leq 0.5 
    \end{aligned}
    \label{eq:sigma_below}
\end{equation}
where $p_\mathrm{O}$ is the pressure at the face centre (outside the interface in this case) and $\kappa \sigma$ corresponds to the Laplacian pressure jump across the interface. In Eq. \ref{eq:sigma_below}, we have assumed that the pressure field stored at the face centre acts on the area of the face outside the interface (i.e., $(1 - \gamma) A$), since the face centre belongs to this region. The remaining part of the face (i.e., $\gamma A$) is inside the interface and the pressure acting on it is computed by enforcing the Laplacian pressure jump. For the case shown in Figure \ref{fig:face_fractions}b, the scheme reads:
\begin{equation}
    \begin{aligned}
        \iint_\mathrm{face} p \,dS\ 
        &\approx (p_\mathrm{O} - \kappa \sigma) (1 - \gamma) A + 
        p_\mathrm{O} \gamma A \\
    &= (p_\mathrm{O} + \kappa \sigma (\gamma - 1)) A
    && \text{for } \gamma > 0.5 
    \end{aligned}
    \label{eq:sigma_above}
\end{equation}
Both cases in Eq. \ref{eq:sigma_below} - \ref{eq:sigma_above} can be rewritten in a unified formulation:
\begin{equation}
    \iint_\mathrm{face} p \,dS\ \approx
    (p_\mathrm{O} + \beta) A
    \label{eq:pressure_scheme}
\end{equation}
where the parameter $\beta$ acts as a correction term for pressure, and is defined as:
\begin{equation}
    \beta = 
    \begin{cases}
        \kappa \sigma \gamma, & \text{if } \gamma \le 0.5 \\
        \kappa \sigma (\gamma - 1), & \text{else}
    \end{cases}
    \label{eq:beta_factor}
\end{equation}
It is finally noted that, for Cartesian grids, the pressure gradient contributes to the $x$--component of the momentum only through the faces with normals $\pm \mathbf{x}$, whereas the surface tension integral in Eq. \ref{eq:surf_tension_integral} must be evaluated on all the faces that bound the control volume. 

By combining the schemes from Eq. \ref{eq:surf_tension_integral} and \ref{eq:pressure_scheme} for surface tension and pressure, respectively, we obtain the total contribution to the $x$--momentum:
\begin{equation}
    - \iiint_\Omega \frac{\partial p}{\partial x} \,dV\ + 
    \oint_{\partial \Sigma} \sigma \mathbf{m}_\Sigma \cdot \mathbf{x} \,dl\ \approx
    -\left\{ \left[(p_\mathrm{O} + \beta) A \right]_{x^+} - \left[(p_\mathrm{O} + \beta) A \right]_{x^-} \right\} + 
    \sum_\mathrm{faces} \left( \sigma l_\Sigma \mathbf{m}_\Sigma \right)_\mathrm{f} \cdot \mathbf{x}
    \label{eq:integral_surf_scheme}
\end{equation}
where $x^+, x^-$ represent the faces with orientation $\mathbf{x}$ and $-\mathbf{x}$, respectively. In order to evaluate the RHS of Eq. \ref{eq:integral_surf_scheme}, the quantities $\beta$, $l_{\Sigma}$ and $\mathbf{m}_{\Sigma}$ must be computed for each face $\mathrm{f}$, the details of which will be explained in the following.

\subsubsection{The front tracking structure and computation of the intersections between the interface and the cells}
\label{sec:The front tracking structure and computation of the intersections between the interface and the cells}
In the front tracking algorithm, the interface between the two phases is represented by an explicit mesh, which typically consists of connected triangles, where the coordinates of the vertices as well as the connectivity among vertices, edges and triangles are stored in dedicated linked lists. 
A detailed description of the FT framework is beyond the scope of this work and the interested reader is referred to the pioneering work of \citet{Tryggvason2001} for a general introduction and to \citep{Gorges2022, Gorges2023, Gorges2025b, Gennari2025} for a detailed description of the FT implementation used in this work.  

In order to compute the geometric properties required to close the scheme of Eq. \ref{eq:integral_surf_scheme} (i.e., $\beta$, $l_{\Sigma}$ and $\mathbf{m}_{\Sigma}$), the intersection between the front mesh representing the interface and the faces of the fluid mesh must be computed. In principle, this operation could be performed directly between the triangulated interface and each cell face. However, such an approach would have to take into account limit relative configurations between the triangles and the face. Such cases include configurations in which one or more triangles or edges lie on the face or pass through one of the four corners. Taking into account all the possible configurations for a three-dimensional interface is not trivial and this approach would lead to an algorithm that lacks the appropriate level of robustness for complex two--phase flow simulations. A further disadvantage would occur for relatively coarse interfaces, where the triangular pattern on the front would not provide a smooth representation of geometric properties, such as the normal vector and curvature, whose accurate computation is essential in surface tension dominated flows.

\begin{figure}[!htbp]
    \centering
    \includegraphics[]{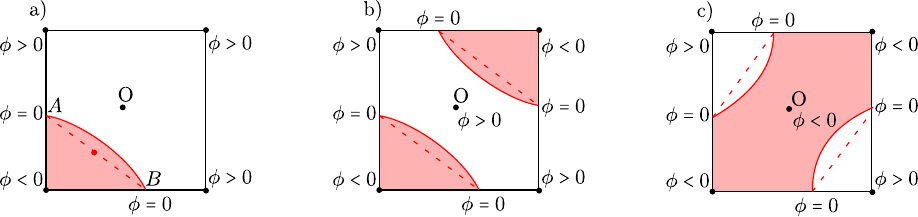}
    \caption{Interface reconstruction from the signed distance field $\phi$ for non--ambiguous (a) and ambiguous (b,c) configurations. For ambiguous cases, the correct solution depends on the distance field value at the centre of the face, i.e., $\phi_\mathrm{O}$. The red area corresponds to the face region inside the interface.}
    \label{fig:signed_distance}
\end{figure}

Given the reasons above, we propose here a different, smoother and more robust, approach to compute the intersection between the interface and the cell faces. The method is based on the signed distance function field $(\phi)$, which is computed for each corner of the face. In this work we assume the signed distance between a point and the interface to be $\phi < 0$ ($\phi > 0$) if the point is inside (outside) the interface. By definition, the interface corresponds to the region where $\phi = 0$. Therefore, once the field $\phi$ is known at the face's corners, the computation of the intersection is straightforward. Figure \ref{fig:signed_distance}a shows one possible configuration in which one corner is inside and the other three are outside. 
The two intersections necessarily occurs along the edges that connect two corners with opposite signs and their location can be computed assuming a linear profile for $\phi$ and looking for the point where $\phi = 0$. An ambiguous configuration can occur if all the four edges are cut by the interface. In this case, two possible configurations are available for the same distribution of $\phi$ at the corners, as shown in Figure \ref{fig:signed_distance}b and \ref{fig:signed_distance}c. In order to chose the right solution, it is necessary to evaluate the distance function at a fifth point (e.g., the face centre $\mathrm{O}$): if $\phi_\mathrm{O} > 0$, the right configuration is necessarily the one represented in Figure \ref{fig:signed_distance}b, otherwise the solution of Figure \ref{fig:signed_distance}c is selected.

In order to compute the signed distance function $\phi$, a generic quadratic surface is fitted through the front markers around each corner of the face. The same surface is also used to compute the normal and the curvature $\kappa$ of the interface, providing accurate predictions of such quantities even at relatively low resolutions. We finally note that a conceptually similar approach is used by \citet{Popinet1999} to compute the tangent direction at the interface, where cubic splines are fitted through the markers of the 2D front.

\begin{figure}[!htbp]
    \centering
    \includegraphics[]{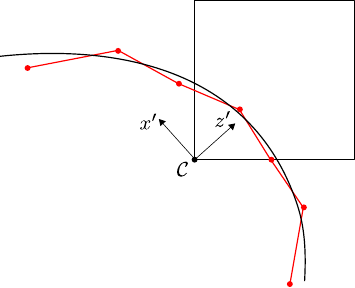}
    \caption{Two--dimensional example of the fitting of a quadratic curve (black line) through the markers of the interface (red line). The local reference system is centred at the face corner $\mathcal{C}$, and the $z'$--direction corresponds to the normal vector from the closest marker.}
    \label{fig:quadratic_fitting}
\end{figure}

The procedure to fit a quadratic surface $\mathcal{Q}(x, y, z)$ through the markers of the interface and compute the relevant geometric properties is summarised in Algorithm \ref{alg:fitting}. A generic quadratic surface reads:  
\begin{equation}
    \mathcal{Q}(x,y) = z = a_1x^2 + a_2xy + a_3y^2 + a_4x + a_5y + a_6
    \label{eq:quadratic}
\end{equation}
and the solution therefore consists of finding the six coefficients $\mathbf{a} = [a_1, a_2, a_3, a_4, a_5, a_6]^\mathrm{T}$ that best approximate the front. This problem is better addressed in a local reference frame $\mathcal{R'}(\mathcal{C}, x', y', z')$ centred on the face corner $\mathcal{C}$, where $\mathbf{z'}$ is the normal vector at the interface computed from the closest front marker; $\mathbf{x', y'}$ are two generic directions that, together with $\mathbf{z'}$, form an orthonormal basis (see Figure \ref{fig:quadratic_fitting} for a 2D example). 
The problem in Eq. \ref{eq:quadratic} is solved in this work via a weighted least--squared approach, which consists of minimising the residual error
\begin{equation}
    \min_{\mathbf{a} \in \mathbb{R}^6} \left[ \sum_{i = 1}^{N_\mathrm{marker}} w_i \left( \mathcal{Q}(x', y') - z' \right)^2 \right]
    \label{eq:least-squares}
\end{equation}
where $N_\mathrm{marker}$ is the number of markers (ranging from 24 to 48) used for the fitting, and $w_i$ is a weight coefficient based on the distance between the $i$--th marker and $\mathcal{C}$ (see Algorithm \ref{alg:fitting} for details on the solution procedure).

Once the problem in Eq. \ref{eq:least-squares} is solved, the distance function and the geometric properties of interest can be found. The normal vector at point $\mathcal{C}$ is:
\begin{equation}
    \mathbf{n}_{\Sigma, \mathcal{C}} = \frac{1}{\sqrt{\left( \frac{\partial \mathcal{Q}}{\partial x'} \right)^2 + \left( \frac{\partial \mathcal{Q}}{\partial y'} \right)^2 + 1}}
    \begin{bmatrix} {\partial \mathcal{Q}}/{\partial x'} \\ {\partial \mathcal{Q}}/{\partial y'} \\ -1 \end{bmatrix} = \frac{1}{\sqrt{a_4^2 + a_5^2 + 1}}
    \begin{bmatrix} a_4 \\ a_5 \\ -1 \end{bmatrix}
    \label{eq:normal_quadratic}
\end{equation}
where the derivatives in Eq. \ref{eq:normal_quadratic} are computed at $\mathcal{C}$, which is the origin of the reference frame $\mathcal{R'}$, i.e., $x'_\mathcal{C} = y'_\mathcal{C} = z'_\mathcal{C} = 0$. The orientation of $\mathbf{n}_{\Sigma, \mathcal{C}}$ is checked against $\mathbf{z}'$ to ensure that the normal consistently points into the same phase (the secondary phase in this work). The curvature computed at $\mathcal{C}$ follows as:
\begin{equation}
    \kappa_\mathcal{C} = - \left( \nabla \cdot  
    \mathbf{n}_\Sigma
    \right)_\mathcal{C} = 
    - \frac{2a_1(1 + a_5^2) - 2a_2a_4a_5 + 2a_3(1 + a_4^2)}{(a_4^2 + a_5^2 + 1)^{3/2}}
    \label{eq:curvature_quadratic}
\end{equation}
The distance between the quadratic surface and point $\mathcal{C}$ along the normal of the reference frame $\mathcal{R'}$, i.e., $\mathbf{z'}$, is simply $-a_6$, where the sign ensures that the signed distance is negative if $\mathcal{C}$ is located inside the interface. However, the $z'$--direction is simply a tentative approximation of the interface normal taken from the closest marker. The actual normal direction of the quadratic surface is given by Eq. \ref{eq:normal_quadratic}, and the signed distance field is computed as:
\begin{equation}
    \phi_\mathcal{C} = - \frac{a_6}{\sqrt{a_4^2 + a_5^2 + 1}}
    \label{eq:signed_distance_quadratic}
\end{equation}
where the correction factor $1/\sqrt{a_4^2 + a_5^2 + 1}$ corresponds to the projection of $\mathbf{z}'$ onto $\mathbf{n}_{\Sigma, \mathcal{C}}$. Such an approximation is accurate when the angle between $\mathbf{z}'$ and $\mathbf{n}_{\Sigma, \mathcal{C}}$ is small and $\mathcal{C}$ is close to the interface. The first requirement is always true, since $\mathbf{z}'$ is the normal at the interface computed from the triangles around the closest marker. The second assumption is not satisfied if the face corner $\mathcal{C}$ is far from the interface. However, in such a scenario an accurate computation of $\phi$ is not required, since all the four corners of the face will be either inside or outside the interface and no intersections occur. An accurate computation of $\phi$ is needed only when the sign of the distance function changes across the four face corners, and this can occur only if the distance between the interface and $\mathcal{C}$ is close or below the mesh size. It is finally noted that computing the distance function by fitting a quadratic surface at each face corner ensures that intersections are consistent when considering two cell faces that share the same edge, i.e., the intersection along that edge is unambiguously defined. 
\begin{algorithm}
\caption{Quadratic fitting and computation of geometric properties}
\label{alg:fitting}
\KwIn{list of front markers $\mathcal{M}$ (with normals), face corner $\mathcal{C}$, mesh size $\Delta$}
\KwOut{Signed distance $\phi$, normal vector $\mathbf{n}_\Sigma$, curvature $\kappa$}
\tcc{Compute local reference frame $\mathcal{R'}(\mathcal{C}, x', y', z')$}
Initialise weight array $w[]$\;
$d_\mathrm{min} = 10^{10}$\;
\ForEach{marker $m \in \mathcal{M}$}{
    Compute the distance between $m$ and $\mathcal{C}$ $(\overline{m \mathcal{C}})$\;
    Compute the marker weight $w[m] \leftarrow e^{- \left( \frac{\overline{m\mathcal{C}}}{\Delta} \right)^2}$\;
    \If {$\overline{m\mathcal{C}} < d_\mathrm{min}$}{
        $d_\mathrm{min} \leftarrow \overline{m\mathcal{C}}$\;
        $m_\mathrm{closest} \leftarrow m$\;
    }
}
Set $\mathbf{z'}$ equal to the normal stored at $m_\mathrm{closest}$\;
Create an orthonormal basis $(\mathbf{x}', \mathbf{y}', \mathbf{z}')$\;
\ForEach{marker $m \in \mathcal{M}$}{
    Transform the coordinates of $m$ into the reference frame $\mathcal{R'}$\;
}
\tcc{Solve the least-square problem for the quadratic fitting}
Initialise system matrix $\mathbf{M}[6][6]$, RHS vector $\mathbf{b}[6]$ and solution vector $\mathbf{a}[6]$\;
\ForEach{marker $m \in \mathcal{M}$}{
    $\mathbf{r} \leftarrow [x_m^2,\, x_m y_m,\, y_m^2,\, x_m,\, y_m,\, 1]$\;
    \For{$\mathit{row} = 0$ \KwTo $5$}{
        \For{$\mathit{col} = \mathit{row}$ \KwTo $5$}{
            $\mathbf{M}[\mathit{row}][\mathit{col}] \mathrel{+}= w[m] \cdot \mathbf{r}[\mathit{row}] \cdot \mathbf{r}[\mathit{col}]$\;
        }
        $\mathbf{b}[\mathit{row}] \mathrel{+}= w[m] \cdot \mathbf{r}[\mathit{row}] \cdot z_m$\;
    }
}
Populate the missing terms of the symmetric matrix $\mathbf{M}$\;
Solve the system $\mathbf{M} \mathbf{a} = \mathbf{b}$\;
\tcc{Compute the geometric properties}
Compute the normal $\mathbf{n}_\Sigma$ (Eq. \ref{eq:normal_quadratic})\;
\If{$(\mathbf{n}_\Sigma \cdot \mathbf{z'} < 0)$}{
    $\mathbf{n}_\Sigma \leftarrow -\mathbf{n}_\Sigma$\;
}
Compute the curvature $\kappa$ (Eq. \ref{eq:curvature_quadratic})\;
Compute the signed distance function $\phi$ (Eq. \ref{eq:signed_distance_quadratic})\;
\Return $\phi$, $\mathbf{n}_\Sigma$, $\kappa$\;
\end{algorithm}

Once $\phi$ is known, the intersections along the face edges (i.e., the points where $\phi = 0$) are computed via linear interpolation. Given two intersection points ($A, B$ in Figure \ref{fig:signed_distance}a), the length of the interface is $l_\Sigma = \overline{AB}$, whereas the tangent direction is $\mathbf{t}_\Sigma = \overrightarrow{AB} / \overline{AB}$. The interfacial normal and curvature are computed by fitting another quadratic surface around the mid--point between $A$ and $B$ (red point in Figure \ref{fig:signed_distance}a), and using Eq. \ref{eq:normal_quadratic} and \ref{eq:curvature_quadratic}, respectively. The surface tension force simply follows as $\sigma l_\Sigma \mathbf{m}_\Sigma = \sigma l_\Sigma (\mathbf{t}_\Sigma \times \mathbf{n}_\Sigma)$. The direction of this force is then checked against the face normal to guarantee that it points in the outward direction with respect to the control volume. 
Once all the intersection points at the face edges are found, the face fraction $\gamma$ (Eq. \ref{eq:gamma_f}) simply follows from the reconstructed interface (dotted line in Figure \ref{fig:signed_distance}a). This term, along with the curvature and $\sigma$ at the mid--point, is used to compute the correction factor $\beta$ (Eq. \ref{eq:beta_factor}).

\subsection{The classic approach in front tracking: a smoothing--based formulation}
\label{sec:The classic approach in front tracking: a smoothing--based formulation}
In this section, we briefly introduce the most common treatment for surface tension in front--tracking frameworks, also referred to as classic FT approach. This scheme will be used as a benchmark in section \ref{sec:Validation}, and it combines properties from both the integral and volumetric formulations. A detailed description is beyond the scope of the work, and the interested reader is referred to \citep{Tryggvason2001, Tryggvason2011a} for more details.

Starting from the integral formulation (Eq. \ref{eq:surf_tension}), the surface tension force acting on a single triangle $T$ (with normal $\mathbf{n}_{\Sigma, T}$) of the front reads:
\begin{equation}
    \oint_{\partial T} \sigma \mathbf{m}_\Sigma \,dl\ \approx 
    \sum_{\mathrm{edges}} \sigma_E \left( \mathbf{t}_{\Sigma, E} \times \mathbf{n}_{\Sigma,E} \right) l_E = \mathbf{F}_{\sigma, T}
    \label{eq:scheme_classic}
\end{equation}
where the summation is performed over each edge $E$ of triangle $T$, and $\mathbf{F}_{\sigma, T}$ is the actual surface tension force acting on $T$. The tangent vector $\mathbf{t}_{\Sigma,E}$ is simply the direction of edge $E$, whereas the unit normal is approximated in this work as $\mathbf{n}_{\Sigma,E} \approx \mathbf{n}_{\Sigma, T}$. Such an approximation is effective since each edge is considered twice (see Eq. \ref{eq:scheme_classic_discretised} below). In a closed front, each edge is shared by two adjacent triangles and when both of them are considered, the resulting normal at the edge corresponds to the average between the normals of the adjacent triangles. The scheme is recast into a volumetric formulation, and the source term $\mathbf{f}_\sigma \delta_\Sigma$ (a volumetric force, see Eq. \ref{eq:surf_tension_vol}) for cell $\Omega$ is obtained from Eq. \ref{eq:scheme_classic} via weighted interpolation:
\begin{equation}
    (\mathbf{f}_{\sigma} \delta_\Sigma)_\Omega = \sum_\mathrm{triangles} 
    \left(
    w_{T, \Omega}  \frac{\mathbf{F}_{\sigma, T}} {V_\Omega} 
    \right)
    \label{eq:scheme_classic_discretised}
\end{equation}
where $V_\Omega$ is the volume of the computational cell. The relative interpolation weight between triangle $T$ and cell $\Omega$ is computed as $w_{T, \Omega} = d(r_x)d(r_y)d(r_z)$, where the smoothing kernel is taken from \citep{Peskin2003, Peskin1972}:
\begin{equation}
\label{eq:peskin}
    d(r)=\begin{cases}
    \frac{1}{4} \left(1 + \mathrm{cos}\left( \frac{1}{2} \pi r \right)\right), & \text{if $|r|<2$}\\
    0, & \text{otherwise}
    \end{cases}
\end{equation}
and $\mathbf{r}$ is the distance between the centres of $\Omega$ and $T$, i.e., $\mathbf{r} = (\mathbf{x}_\Omega - \mathbf{x}_T)/\Delta$, made non--dimensional with the cell size $\Delta$. The smoothing kernel in Eq. \ref{eq:peskin} ensures that the contribution from each triangle is smoothly distributed only to the surrounding fluid cells of the Eulerian mesh. 

It is noted that, having started from an integral formulation and later moved to a volumetric description, the classic FT scheme shares properties of both approaches. It does not require the explicit computation of curvature and works naturally for cases characterised by variable surface tension. However, due to the spreading into a volumetric source term, it does not conserve momentum neither locally nor globally. 

\subsection{The continuous surface force scheme}
\label{sec:The continuous surface force scheme}
One of the most popular surface tension treatments in two--phase flow simulations is the continuous surface force (CSF) scheme, originally proposed by \citet{Brackbill1992}, which belongs to the family of volumetric approaches. This scheme, along with the classic FT one, will be used in the following as a benchmark for the integral approach discussed in section \ref{sec:The integral surface tension scheme and its implementation into a front tracking framework}. A detailed discussion of the CSF scheme is beyond the scope of this work, and the interested reader is referred to \citep{Popinet2018} for a detailed discussion on its properties. In the following, CSF will be used only for cases with constant surface tension coefficients, therefore we introduce it by here assuming $\mathbf{f}_\sigma = \sigma \kappa \mathbf{n}_\Sigma$, i.e., $\nabla_\Sigma \sigma = 0$.

By introducing the Heaviside function $H$ (which has values $1$ and $0$ in the primary and secondary phases, respectively) and exploiting the definition of $\delta_\Sigma$ (i.e., $\delta_\Sigma \mathbf{n}_\Sigma = \nabla H$ \citep{Tryggvason2011a}), the volumetric surface tension term is rewritten as:
\begin{equation}
    \sigma \kappa \mathbf{n}_\Sigma \delta_\Sigma = \sigma \kappa \nabla H
    \label{eq:vol_surf_tension_H}
\end{equation}
where the interface is located at $\mathbf{x} = \mathbf{x}_\Sigma$. For the discretisation of Eq. \ref{eq:vol_surf_tension_H}, a numerical approximation of the gradient of the Heaviside function is required. In the CSF method, $H$ is approximated with the volume fraction field $\alpha$, and the scheme applied to a control volume $\Omega$ reads as:
\begin{equation}
    \mathbf{f}_{\sigma} \delta_\Sigma \approx (\sigma \kappa \nabla \alpha )_\Omega
    \label{eq:CSF}
\end{equation}
where the subscript $\Omega$ refers to an appropriate numerical discretisation of curvature and gradient of volume fraction for cell $\Omega$. In the present front tracking framework, the volume fraction field is computed via a piecewise-parabolic interface calculation (PPIC) approach, which consists of fitting for each cell $\Omega$ a quadratic surface that approximates the interface, using an algorithm similar to the one discussed in section \ref{sec:The front tracking structure and computation of the intersections between the interface and the cells}. The resulting least--squares problem aims at minimising the volume between the quadratic surface and the local triangles of the front mesh. For a detailed discussion on the PPIC implementation, the reader is referred to the work of \citet{Gorges2025b}. Once the intersection volume $V_{\mathcal{Q}, \Omega}$ between the quadratic surface $\mathcal{Q}$ and cell $\Omega$ is computed, the volume fraction immediately follows as:
\begin{equation}
    \alpha_\Omega = \frac{V_{\mathcal{Q}, \Omega}}{V_\Omega}
    \label{eq:PPIC}
\end{equation}
whereas the curvature $\kappa_\Omega$ is set equal to the curvature of the quadratic surface $\mathcal{Q}$.

Compared to the classic FT scheme of Eq. \ref{eq:scheme_classic_discretised}, CSF has the disadvantages that an explicit computation of the curvature is required (where other approaches than PPIC are obviously possible, such as height functions \citep{Popinet2009, Evrard2020}) and the extension to variable surface tension problems is non--trivial. Despite these drawbacks, CSF is one of the most employed schemes in the literature. The reason behind its popularity lies in the fact that if offers a natural way to ensure a well--balanced implementation. Well--balanced methods are able to recover equilibrium solutions that typically arise from the balancing between pressure and surface tension (e.g., bubble/droplet in a static flow). Since both therms appear as gradients in the momentum equation ($\nabla p$ and $\nabla \alpha$, respectively), it is straightforward to use the same numerical discretisation for both gradients to reach a well--balanced and consistent implementation of surface tension \citep{Popinet2018}. Obviously, this condition alone is not sufficient to remove spurious velocities, since an accurate approximation of the curvature is likewise crucial.

\section{Governing equations and flow solver}
\label{sec:Governing equations and flow solver}
The system of governing equations for the conservation of mass and momentum of an incompressible two--phase flow is given by the Navier--Stokes equations, which are solved in this work in the one--fluid formulation:
\begin{align}
    \label{eq:continuity}
    \nabla \cdot \mathbf{u} 
    &= 0 \\
    \label{eq:momentum}
    \rho \left( \frac{\partial \mathbf{u}}{\partial t} + \nabla \cdot ( \mathbf{u} \otimes \mathbf{u}) \right) 
    &= - \nabla p + \nabla \cdot (2\mu\mathbf{D}) + \rho \mathbf{g} + \mathbf{S}
\end{align}
where $\mathbf{u}$ is the fluid velocity field, $p$ is the pressure field, $\rho$, $\mu$ are the phase density and dynamic viscosity, respectively, $\mathbf{g}$ is the gravitational acceleration, and the deformation tensor is $\mathbf{D} = \left[ \nabla \mathbf{u} +  \nabla \mathbf{u} ^T \right]/2$. In the present work, the generic source term $\mathbf{S}$ reduces to surface tension only, i.e., $\mathbf{S} = \mathbf{f}_\sigma \delta_\Sigma$. The flow solver is based on a coupled, pressure--based, finite--volume framework, which is second--order accurate in space and time \citep{Denner2014b, Denner2020}. The primary variables are stored at cell centres in a collocated arrangement, and the coupling between pressure and velocity is enforced via a momentum--weighted interpolation (MWI) approach \citep{Bartholomew2018}. The fluid properties $\rho, \mu$ are discontinuous at the interface, since the two phases have typically different densities and viscosities. In the one--fluid formulation, a single set of governing equations is solved across the whole domain, and the fluid properties in a generic cell $\Omega$ are computed as:
\begin{align}
    \rho_\Omega(\mathbf{x}) &= \alpha_\Omega \rho_\mathrm{d} + (1 - \alpha_\Omega) \rho_\mathrm{c}\\
    \mu_\Omega(\mathbf{x}) &= \alpha_\Omega \mu_\mathrm{d} + (1 - \alpha_\Omega) \mu_\mathrm{c}
\end{align}
where the subscripts (d,c) refer to the disperse (primary) and continuous (secondary) phases, respectively. The volume fraction field $\alpha$ can be computed in two ways, depending on the selected surface tension scheme. For the classic approach (Eq. \ref{eq:scheme_classic_discretised}), the volume fraction is spread on the Eulerian mesh, i.e., the mesh where Eq. \ref{eq:continuity} - \ref{eq:momentum} are solved, by solving a Poisson equation with the same smoothing kernel adopted for the surface tension term (Eq. \ref{eq:peskin}) \citep{Gorges2022}. This choice ensures consistency between the volume fraction and surface tension fields in the classic FT framework. On the other hand, the integral surface tension and CSF implementations (Eq. \ref{eq:integral_surf_scheme} and \ref{eq:CSF}, respectively) rely on a sharp representation of the volume fraction field, which is obtained from Eq. \ref{eq:PPIC} via a PPIC reconstruction of the front \citep{Gorges2025b}.

The interface is represented by a triangular front mesh, which is advected at every time step in a Lagrangian fashion:
\begin{equation}
    \frac{\mathrm{d} \mathbf{x}_i}{\mathrm{d} t} = \mathbf{\overline{u}}(\mathbf{x}_i)
    \label{eq:front_advection}
\end{equation}
where $\mathbf{x}_i$ is the position of the $i$--th marker. A divergence--preserving interpolation scheme \citep{Gorges2022} is adopted to interpolate the velocity field at the marker locations (i.e., $\mathbf{\overline{u}}(\mathbf{x}_i)$), whereas a normal--only advection method \citep{Gorges2023} is used to solve Eq. \ref{eq:front_advection}.

\subsection{Well--balanced implementation of surface tension}
As anticipated in section \ref{sec:The continuous surface force scheme}, a well--balanced implementation of the surface tension term (or any other source term that generates a pressure gradient) is essential to recover equilibrium solutions. Since the numerical solver used in this work is based on a collocated arrangement and coupled solution of the governing equations, where pressure--velocity coupling is enforced through the MWI technique, it is essential to build a cell--centred source term $(\mathbf{S}^*)$ that matches the discretisation of $\nabla p$ to achieve a well--balanced implementation. Here we briefly introduce the scheme used in this work to compute $\mathbf{S}^*$; for a detailed derivation the reader is referred to the work of \citet{Bartholomew2018}. Once the $\mathbf{S}^*$ term is computed, it is applied to both the RHS of the momentum equation (where it replaces $\mathbf{S}$ in Eq. \ref{eq:momentum}) and the MWI formulation for the velocity fluxes.

\begin{figure}[!htbp]
    \centering
    \includegraphics[]{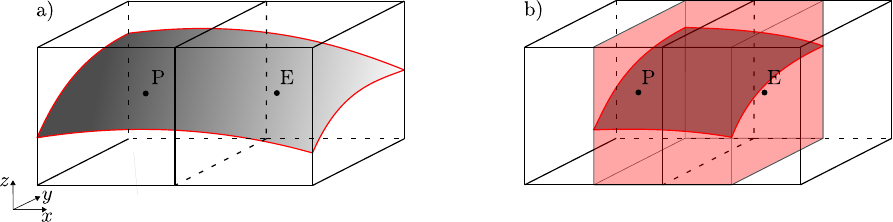}
    \caption{The implementation of well--balanced cell--centred source terms requires, first, the computation of the source term at the cell faces. (a) The CSF method accomplishes this by computing the volume fraction gradient from the values stored at the cell centres and interpolating the curvature at the face. (b) The integral method computes the surface tension force on a staggered volume centred on the same face (red volume).}
    \label{fig:well_balanced}
\end{figure}

Assuming a perfect equilibrium between pressure gradient and source terms in a quiescent flow, the momentum equation in the $x$--direction reads:
\begin{equation}
    0 = -\frac{\partial p}{\partial x} + S_x
    \label{eq:pressure_grad_source_term}
\end{equation}
The discretisation of the cell-centred pressure gradient $P$ (Figure \ref{fig:well_balanced}) is:
\begin{equation}
    \left. \frac{\partial p}{\partial x} \right|_\mathrm{P} \approx 
    \frac{1}{V_\mathrm{P}} \sum_{\mathrm{faces}} p_\mathrm{f} \left( \mathbf{n} \cdot \mathbf{x}  \right) A
    \label{eq:pressure_gradient}
\end{equation}
where P is the cell centre, whereas $p_\mathrm{f}$ and $\mathbf{n}$ are the pressure and outward normal, respectively, at the face with area $A$. 
In the general case of non--uniform grids, the pressure at each cell face is obtained from the cell-centred values as:
\begin{equation}
    p_\mathrm{f} = l_\mathrm{f}p_\mathrm{P} + \left(1 - l_\mathrm{f} \right)p_\mathrm{E}
    \label{eq:pressure_face}
\end{equation}
where $l_\mathrm{f}$ is an interpolation coefficient and E is the centre of the neighbour cell; for equidistant Cartesian grids $l_\mathrm{f} = 1/2$. Inserting Eq. \ref{eq:pressure_face} into Eq. \ref{eq:pressure_gradient} and rearranging, we obtain:
\begin{equation}
    \begin{aligned}
        \frac{1}{V_\mathrm{P}} \sum_{\mathrm{faces}} p_\mathrm{f} \left( \mathbf{n} \cdot \mathbf{x}  \right) A
        &=
        \frac{1}{V_\mathrm{P}} \sum_{\mathrm{faces}} \left( l_\mathrm{f}p_\mathrm{P} + (1 - l_\mathrm{f}) p_\mathrm{E}\right) n_{x} A - \frac{1}{V_\mathrm{P}} \cancelto{0}{\sum_\mathrm{faces} p_\mathrm{P} n_{x} A} \\
        &=
        \frac{1}{V_\mathrm{P}} \sum_\mathrm{faces} 
        \underbrace{
        (1 - l_\mathrm{f})  n_{x} A 
        }_{w_{\mathrm{f}, x}}
        \Delta p_\mathrm{f} \\
        &=
        \frac{1}{V_\mathrm{P}} \sum_\mathrm{faces} w_{\mathrm{f}, x} \Delta p_\mathrm{f}
    \end{aligned}
    \label{eq:pressure_gradient_scheme}
\end{equation}
where $\Delta p_\mathrm{f} = p_\mathrm{E} - p_\mathrm{P}$. From Eq. \ref{eq:pressure_grad_source_term}, it follows that a source term results in a contribution to the pressure gradient. Such a balance can be written across face f as:
\begin{equation}
    \frac{\Delta p_\mathrm{f}}{\Delta s_\mathrm{f}} = \mathbf{S}_\mathrm{f} \cdot \mathbf{n} 
    \label{eq:pressure_source_term_face}
\end{equation}
where $\mathbf{S}_\mathrm{f}$ is the face--centred source term, and $\Delta s_\mathrm{f}$ is the distance between the two cell centres that share face f (e.g., $\overline{PE}$ in Figure \ref{fig:well_balanced}). By substituting Eq. \ref{eq:pressure_source_term_face} into Eq. \ref{eq:pressure_gradient_scheme}, we derive the expression for a cell--centred source term $\mathbf{S}^*$ that matches the discretisation of the pressure gradient:
\begin{equation}
    S^*_{\mathrm{P}, x} = \frac{1}{V_\mathrm{P}} \sum_\mathrm{faces} w_{\mathrm{f}, x} S_{\mathrm{f},x} \Delta s_\mathrm{f}
    \label{eq:S*}
\end{equation}
From Eq. \ref{eq:S*}, it follows that the last step required to build the cell--centred $\mathbf{S}^*$ field is the computation of the face--centred source term $\mathbf{S}_\mathrm{f}$. For the CSF method on an equidistant Cartesian mesh with size $\Delta$, $\mathbf{S}_\mathrm{f}$ is evaluated as ($x$--component, Figure \ref{fig:well_balanced}a):
\begin{equation}
    S_{\mathrm{f}, x}^{\mathrm{CSF}}  
    =
    \sigma 
    \left( \frac{\kappa_\mathrm{P} + \kappa_\mathrm{E}}{2} \right)
    \frac{\alpha_\mathrm{E} - \alpha_\mathrm{P}}{\Delta}
    \label{eq:S_f_CSF}
\end{equation}
where $\kappa_\mathrm{P}, \kappa_\mathrm{E}$ $(\alpha_\mathrm{P}, \alpha_\mathrm{E})$ are the curvatures (volume fractions) computed at the cells with centre $\mathrm{P}$ and $\mathrm{E}$, respectively (see Figure \ref{fig:well_balanced}a).

For the integral formulation, the face--centred source term is computed on a staggered volume centred on face f (the red volume in the example of Figure \ref{fig:well_balanced}b). Rearranging the integral surface tension scheme (Eq. \ref{eq:integral_surf_scheme}) for an equidistant Cartesian mesh leads to:
\begin{equation}
    - \iiint \frac{\partial p}{\partial x} \,dV\ + 
    \oint_{\partial \Sigma} \sigma \mathbf{m}_\Sigma \cdot \mathbf{x} \,dl\ \approx
    -(p_\mathrm{E} - p_\mathrm{P})\Delta^2 + 
    \underbrace{
    \sum_\mathrm{faces} \left( \sigma l_\Sigma \mathbf{m}_\Sigma \right)_\mathrm{f} \cdot \mathbf{x} - 
    (\beta_\mathrm{E} - \beta_\mathrm{P}) \Delta^2
    }_{S_{\mathrm{f},x} \Delta^3}
\end{equation}
where the first term on the RHS is the usual discretisation of the face--centred pressure gradient, which depends on the available data stored at the cell centres; the coefficient that multiplies the pressure values is directly added to the system matrix for the coupled resolution of the governing equations. The last two terms on the RHS depend entirely on the front (and its intersection with the Eulerian grid), and do not contain any of the primary variables; the pressure correction factors $\beta_\mathrm{P}, \beta_\mathrm{E}$ are computed at the $x$--oriented faces passing through the cell centres $\mathrm{P}$ and $\mathrm{E}$, respectively (see Figure \ref{fig:well_balanced}b). These terms are computed explicitly at the beginning of each time step (after the advection of the interface) and their contribution is added to the RHS of the Navier--Stokes system. The face--centred source term, therefore, reads:
\begin{equation}
    S_{_{\mathrm{f},x}}^{\mathrm{integral}} 
     = 
    \frac{1}{\Delta^3} 
    \left[
    \sum_\mathrm{faces} \left( \sigma l_\Sigma \mathbf{m}_\Sigma \right)_\mathrm{f} \cdot \mathbf{x} - 
    (\beta_\mathrm{E} - \beta_\mathrm{P}) \Delta^2
    \right]
\end{equation}
It is noted that for the $y$-- and $z$--components (i.e., $S^\mathrm{integral}_{\mathrm{f},y}$ and $S^\mathrm{integral}_{\mathrm{f},z}$), the pressure correction factor $\beta$ is evaluated at faces passing through cell centres and oriented along the $y$-- and $z$--directions, respectively. In order to ensure consistency of $\beta$ across the three components of the source term, the condition $\gamma \gtreqless 0.5$ is replaced in Eq. \ref{eq:beta_factor} by an equivalent expression for the volume fraction field, i.e., $\alpha \gtreqless 0.5$, which guarantees consistency across all the directions, since it is a volume--based quantity stored at cell centres, and does not depend on the orientation of the faces. 

\begin{figure}[!htbp]
    \centering
    \includegraphics[]{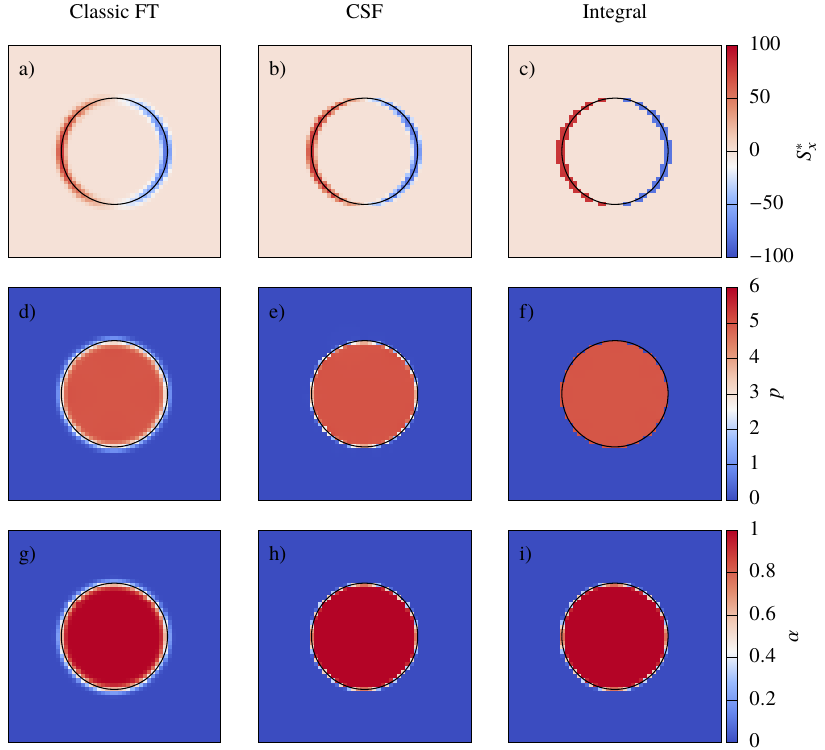}
    \caption{Comparison of the classic FT, CSF and integral surface tension scheme in terms of surface tension distribution ($x$--component) (a--c), pressure  (d--f) and volume fraction (g--i) fields for a spherical droplet in static equilibrium.}
    \label{fig:La12000_all_static_contour_surftension_p_I}
\end{figure}

A qualitative comparison between the three surface tension schemes adopted in this work is shown in Figure \ref{fig:La12000_all_static_contour_surftension_p_I} for a static circular droplet. The classic FT and CSF approaches result in a smoother distribution of the source term $\mathbf{S}^*$ around the interface, whereas a sharper representation is achieved by the integral model (Figures \ref{fig:La12000_all_static_contour_surftension_p_I}a--c). The pressure fields shown in Figures \ref{fig:La12000_all_static_contour_surftension_p_I}d--f mimic the distribution of the volume fraction fields (Figure \ref{fig:La12000_all_static_contour_surftension_p_I}g--i) for the classic and CSF approaches. In the classic FT framework, the volume fraction field is smoothed across 3--4 layers of cells and so is the pressure, whereas in the CSF scheme $p$ jumps between $p_\mathrm{in}$ and $p_\mathrm{out}$ across one cell only. On the other hand, the correction factor $\beta$ of the integral scheme, which explicitly takes into account the static pressure jump across the interface, forces every cell with $\alpha \leq 0.5$ $(\alpha > 0.5)$ to take the value $p = p_\mathrm{out}$ $(p = p_\mathrm{in})$, resulting in a very accurate representation of such a discontinuous field.

\section{Validation}
\label{sec:Validation}
In this section, the proposed integral surface tension scheme is validated across several benchmarks, with a focus on equilibrium solutions and surface--tension driven flows. The numerical framework is tested against classic FT and CSF approaches, and the results are compared against analytical solutions and/or experimental data. Many equilibrium solutions depend on the exact balance between pressure gradient and surface tension force. Remeshing operations, required in FT simulations to regularise the front mesh \citep{Gorges2022}, typically introduce perturbations that alter this balance. Our previous results on the modelling of surface tension dominated flows with FT \citep{Gorges2025b} have shown that accuracy can be significantly enhanced by adopting a remeshing technique based on a localised roughness smoothing approach (see \citep{Gorges2025b} for details). This technique, however, requires the fine tuning of specific parameters that are typically configuration-dependent. At the cost of some accuracy, we have decided to rely only on well established remeshing techniques, such as edge splitting, merging and flipping, as well as global smoothing operations, such as tangential relaxation \citep{Botsch2011} and TSUR--3D \citep{deSousa2004}.   

In the following, the subscript (d,c) will be used to refer to the properties of the disperse (i.e., bubble/droplet) and continuous (i.e., ambient) phases, respectively. The terms Eulerian or fluid mesh will be used to refer to the mesh where the governing equations are solved, whereas Lagrangian mesh or front refer to the front tracking interface.

\subsection{Spurious velocities around a static spherical droplet}
\label{sec:Spurious velocities around a static spherical droplet}
The modelling of a spherical droplet in a quiescent flows is a widely--used benchmark for surface--tension models in two--phase flow simulations \citep{Gorges2025b, Abu-Al-Saud2018, Abadie2015, Popinet2009, Popinet1999, Shin2005, Renardy2002}. This problem consists of a spherical bubble/droplet placed at the centre of the computational domain, with initial velocity field $\mathbf{u}_\mathrm{init} = \mathbf{0}$ and null gravity. With a constant surface tension coefficient $\sigma$, this configuration results in the equilibrium solution:
\begin{equation}
\label{eq:young_laplace}
    \begin{aligned}
        \Delta p = p_\mathrm{in} - p_\mathrm{out} &= \sigma \kappa \\
        \mathbf{u}(\mathbf{x}, t) &= \mathbf{0}
    \end{aligned}
\end{equation}
where the mean curvature for a sphere is $\kappa = 2/R$, and $R$ is the radius. The first relationship in Eq. \ref{eq:young_laplace} is the well--known Young--Laplace equation. Due to numerical errors introduced by discretisation schemes, solver tolerances and, most importantly, inaccuracies in the computation of the mean curvature $\kappa$, Eq. \ref{eq:young_laplace} is typically not satisfied exactly, leading to an imbalance between pressure gradient and surface tension force. As a result, spurious velocities are generated around the interface. This numerical artefact is used to quantify the accuracy of the numerical framework against the analytical solution. Well--balanced schemes have proven capable of reducing spurious velocities down to machine accuracy in a viscous time unit. However, these results are achieved in a 2D framework and employ symmetry planes to reduced the extension of the computational domain, see for example \citep{Popinet2009, Abadie2015}. 

In this work, a sphere with $D = 0.8$ is placed at the centre of a cubic domain with size $2.5D \times 2.5D \times 2.5D$. The numerical setup is adopted from \citet{Gorges2025b}, where the same FT framework, with classic and CSF surface tension schemes, is employed. A symmetry boundary condition is applied at the walls of the domain, but since a full 3D interface is modelled, the sphere is not constrained to a fixed position. The problem is formulated in a non--dimensional form, and fully described by the following density and viscosity ratios:
\begin{equation}
    \rho_\mathrm{r} = \frac{\rho_\mathrm{d}}{\rho_\mathrm{c}} = 1, \quad\ \mu_\mathrm{r} = \frac{\mu_\mathrm{d}}{\mu_\mathrm{c}} = 1,
\end{equation}
and Laplace number:
\begin{equation}
    \mathrm{La} = \frac{D \sigma \rho_\mathrm{c}}{\mu_\mathrm{c}^2} = 12000
    \label{eq:laplace_number}
\end{equation}
A further non--dimensional number, the capillary number, is used to quantify the spurious velocities. Two definitions of the capillary number will be used in the following, namely:
\begin{equation}
    \mathrm{Ca_{max}} = \frac{\mu_\mathrm{c} u_\mathrm{max}}{\sigma} ,\quad\ 
    \mathrm{Ca_{rms}} = \frac{\mu_\mathrm{c} u_\mathrm{rms}}{\sigma}
    \label{eq:Camax_Carms}
\end{equation}
where the maximum $(u_\mathrm{max})$ and root mean square $(u_\mathrm{rms})$ velocities are computed, respectively, as:
\begin{equation}
    u_\mathrm{max} = \max_{i \in 1, \dots, N} |\mathbf{u}_i| ,\quad\
    u_\mathrm{rms} = \sqrt{\frac{1}{V} \sum_{i=1}^N |\mathbf{u}_i|^2 V_i}
\end{equation}
where $i$ is the index of the $i$--th cell with volume $V_i$, whereas $V$ is the total volume of the computational domain, i.e., $V = \sum_{i=1}^N V_i$; the total count of cells is $N$. The time step is constrained by the capillary time step restriction \citep{Denner2015a}. 

\begin{figure}[!htbp]
    \centering
    \includegraphics[]{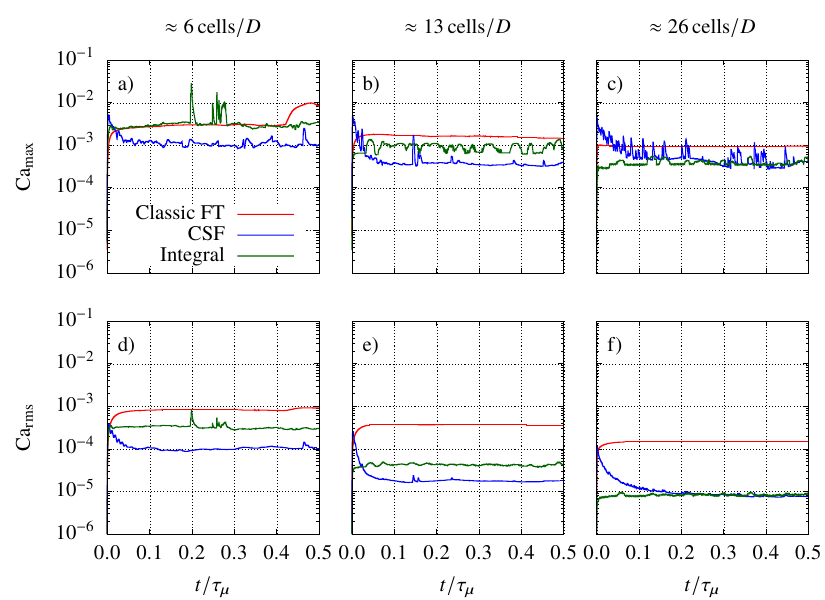}
    \caption{Comparison of time--evolving capillary numbers based on maximum (a--c) and root mean square (d--f) velocities for a static droplet $(\mathrm{La = 12000})$ at three different mesh resolutions.}
    \label{fig:La12000_all_static}
\end{figure}

The results for the time evolution of the capillary numbers for classic FT, CSF and integral surface tension schemes are shown in Figure \ref{fig:La12000_all_static}, for three different mesh resolutions, namely 6, 13 and 26 cells per droplet diameter. Time is made non-dimensional with the viscous time unit $\tau_\mu = \rho_\mathrm{c} D^2/\mu_\mathrm{c}$.
For the low and intermediate resolutions, CSF outperforms classic and integral schemes, resulting in the lowest $\mathrm{Ca_{max}}$ and $\mathrm{Ca_{rms}}$. The integral scheme sits between CSF and classic approaches, whereas the classic approach results in the poorest model, consistent with our previous results \citep{Gorges2025b}. At the lowest resolution (Figure \ref{fig:La12000_all_static}a), the integral scheme exhibits strong fluctuations between approximately $0.2 < t/\tau_\mu < 0.3$. At such a low resolution, the smoothing of the front can induce a non--negligible difference in the shape of the quadratic surface fitted through the front markers, which, in turn, results in a sudden change of the volume fraction field obtained via the PPIC reconstruction (Eq. \ref{eq:PPIC}), e.g., from $\alpha < 0.5$ to $\alpha > 0.5$, or vice versa. As a consequence, the pressure correction factor $\beta$ (Eq. \ref{eq:beta_factor}) abruptly changes sign between two consecutive time steps, forcing the pressure field to jump from $p_\mathrm{out}$ to $p_\mathrm{in}$, or the other way around. When this event occurs, a local perturbation is introduced in the balance between pressure gradient and surface tension, which results in the production of flow momentum. Viscous dissipation will act on such a perturbation until the system approaches a new equilibrium state, which occurs for $t/\tau_\mu > 0.3$ in Figure \ref{fig:La12000_all_static}a. It is noted that such a low mesh refinement is typically not enough to resolve common interfacial flows. With the more practical resolutions of 13 and 26 cells per diameter, sharp fluctuations of the integral scheme are never observed.  

\begin{figure}[!htbp]
    \centering
    \includegraphics[]{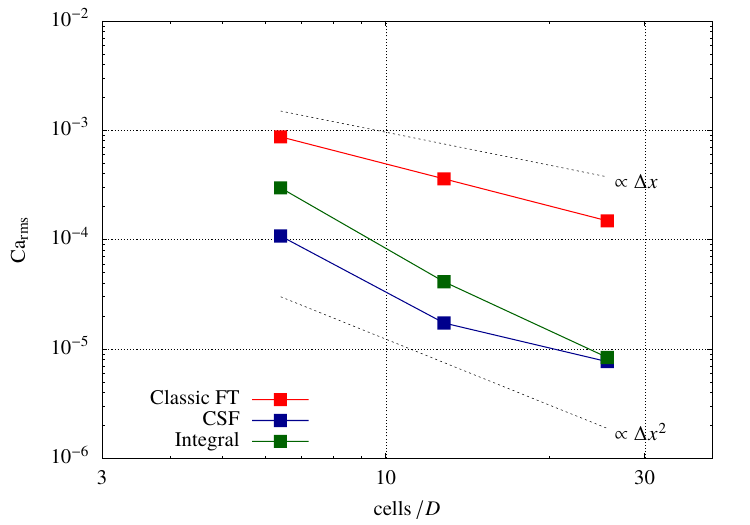}
    \caption{Analysis of the convergence rate of the root mean square capillary number for a static droplet $(\mathrm{La = 12000})$, employing the classic FT, CSF and integral surface tension schemes. Results are averaged for $0.4 < t/\tau_\mu < 0.5$.}
    \label{fig:La12000_all_static_convergence}
\end{figure}

At the highest resolution, the integral scheme reaches the same rms level of spurious velocities as the CSF method (Figure \ref{fig:La12000_all_static_convergence}f), but has the advantage of exhibiting lower fluctuations in the maximum value of the capillary number (Figure \ref{fig:La12000_all_static_convergence}c). Differences between these two schemes also include the trend of the numerical artefacts. The CSF approach typically starts from larger velocities, until viscous dissipation drives the system towards the equilibrium solution, which occurs at $t/\tau_\mu \approx 0.1$. On the other hand, the integral scheme reaches almost immediately the steady--state solution and the capillary numbers remain constant throughout the simulation.

\begin{figure}[!htbp]
    \centering
    \includegraphics[]{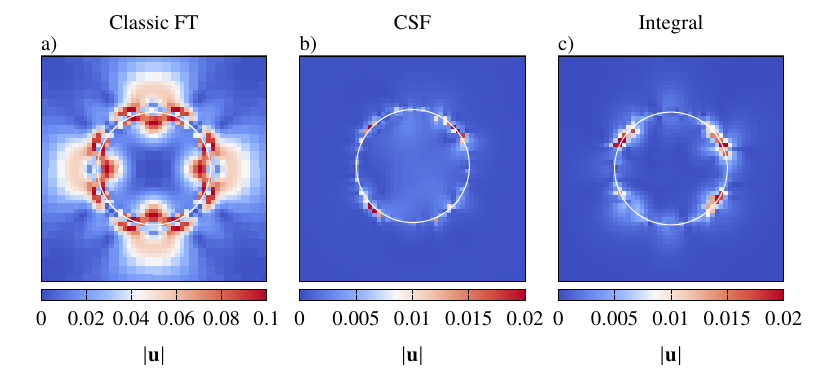}
    \caption{Contours of velocity magnitude on the $x$--$y$ plane for a static droplet $(\mathrm{La = 12000})$ at $t/\tau_\mu \approx 0.4$, for the classic FT (a), CSF (b) and Integral (c) schemes.}
    \label{fig:La12000_all_static_contour}
\end{figure}

A qualitative behaviour of the different surface tension treatments is shown in Figure \ref{fig:La12000_all_static_contour}, where the contours of the velocity magnitude are compared for the finest resolution (i.e., $\approx26$ cells$/D$).
The classic FT scheme produces large regions of non--null velocities, whereas the CSF and integral solutions result in localised (around the interface only) zones of spurious velocities, consistent with the plots shown in Figure \ref{fig:La12000_all_static}f, where the $\mathrm{Ca_{rms}}$ is more than one order of magnitude smaller than classic FT.  

The convergence rate for the three schemes is reported for $\mathrm{Ca_{rms}}$ in Figure \ref{fig:La12000_all_static_convergence}, which shows that classic and integral converge with first-- and second--order accuracy, respectively. The CSF method sits in between, exhibiting second--order behaviour between the coarse and medium resolutions, but only first--order towards the finest grid. 

\subsection{Spurious velocities around a translating spherical droplet}
\label{sec:Spurious velocities around a translating spherical droplet}
This test case is built on the static droplet one and aims at quantifying spurious velocities in a dynamic configuration. A spherical droplet with $D=0.4$ is placed in a domain with size $12.5D \times 2.5D \times 2.5D$ and initial position $\mathbf{x}_\mathrm{d}(t=0) = [D, 0, 0]$. A uniform velocity field $(\mathbf{U}_\mathrm{ref})$ is initialised everywhere as $\mathbf{U}_\mathrm{ref} = [1,0,0]$; inlet and outlet boundary conditions are applied at the domain boundaries $x=0$ and $x = 12.5D$, whereas symmetric conditions are used for the other walls. Since the flow field has uniform velocity $\mathbf{u}(\mathbf{x}, t) = \mathbf{U}_\mathrm{ref}$, the solution (in a reference frame moving with the droplet) is the same as the static droplet test case (Eq. \ref{eq:young_laplace}). In the domain reference frame, the solution simply reads:
\begin{equation}
\label{eq:young_laplace_moving}
    \begin{aligned}
        \Delta p = p_\mathrm{in} - p_\mathrm{out} &= \sigma \kappa \\
        \mathbf{u}(\mathbf{x}, t) &= \mathbf{U}_\mathrm{ref}
    \end{aligned}
\end{equation}
Since a new independent variable has been introduced (i.e., $U_{\mathrm{ref},x},$), an additional non--dimensional number is required to define the problem. The present test case is characterised by the following density and viscosity ratios:
\begin{equation}
    \rho_\mathrm{r} = \frac{\rho_\mathrm{d}}{\rho_\mathrm{c}} = 1, \quad\ \mu_\mathrm{r} = \frac{\mu_\mathrm{d}}{\mu_\mathrm{c}} = 1,
\end{equation}
Laplace number:
\begin{equation}
    \mathrm{La} = \frac{D \sigma \rho_\mathrm{c}}{\mu_\mathrm{c}^2} = 12000,
\end{equation}
and Weber number:
\begin{equation}
    \mathrm{We} = \frac{\rho_\mathrm{c} U_{\mathrm{ref},x}^2 D}{\sigma} = 0.4
\end{equation}

Although from a physical point of view this problem is exactly equivalent to the corresponding static one (Galilean invariance), from a numerical perspective it adds an extra layer of complexity. The interface moves continuously, and numerical errors due to velocity interpolation at markers' locations and prediction of their displacements (Eq. \ref{eq:front_advection}) are introduced at every time step. Therefore, errors in the advection of the interface affect directly the surface tension term, which, in turn, produces velocity fluctuations around the interface. To the best of the authors' knowledge, not even 2D simulations with symmetric boundary conditions can recover machine accuracy for spurious velocity under these circumstances. Along with the static droplet example, this test case is a widely adopted benchmark for surface tension and interface advection schemes \citep{Popinet1999, Saini2025a, Abu-Al-Saud2018, Abadie2015, Gorges2025b}.

\begin{figure}[!htbp]
    \centering
    \includegraphics[]{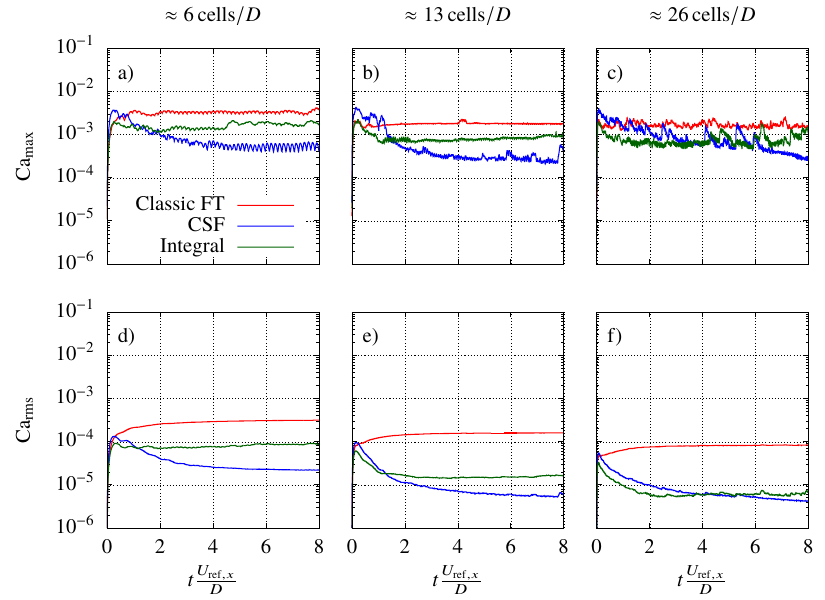}
    \caption{Comparison of time--evolving capillary numbers based on maximum (a--c) and root mean square (d--f) velocities for a translating droplet ($\mathrm{La = 12000}$, $\mathrm{We} = 0.4$) at three different mesh resolutions. Results are shown in the droplet reference frame.}
    \label{fig:La12000_all_translating}
\end{figure}
Spurious velocities (in terms of $\mathrm{Ca_{max}}$ and $\mathrm{Ca_{rms}}$, Eq. \ref{eq:Camax_Carms}) are compared in Figure \ref{fig:La12000_all_translating} for three different mesh resolutions, namely 6, 13 and 26 cells per droplet diameter. The capillary numbers are based on the corresponding velocities in the droplet reference frame, i.e.,:
\begin{equation}
    u_\mathrm{max} = \max_{i \in 1, \dots, N} |\mathbf{u}_i - \mathbf{U}_\mathrm{ref}| ,\quad\
    u_\mathrm{rms} = \sqrt{\frac{1}{V} \sum_{i=1}^N |\mathbf{u}_i - \mathbf{U}_\mathrm{ref}|^2 V_i}
\end{equation}
Each simulation is run until $t = 8D/{U_{\mathrm{ref},x}}$, which corresponds to a translation of 8 droplet diameters. Results are qualitatively analogous to the corresponding static benchmark, with the CSF scheme outperforming classic FT and integral approaches for low and intermediate resolutions, followed by the integral scheme, whereas classic FT always results in higher spurious velocities. At the largest resolution (Figure \ref{fig:La12000_all_translating}c and \ref{fig:La12000_all_translating}f), however, CSF and integral surface tension schemes have similar $\mathrm{Ca_{rms}}$ profiles. It is also worth noting that for the first half of the simulation ($t < 4D/{U_{\mathrm{ref},x}}$), the integral approach provides the lowest spurious velocities, before reaching slightly larger values (compared to CSF) in the steady--state regime. 

\begin{figure}[!htbp]
    \centering
    \includegraphics[]{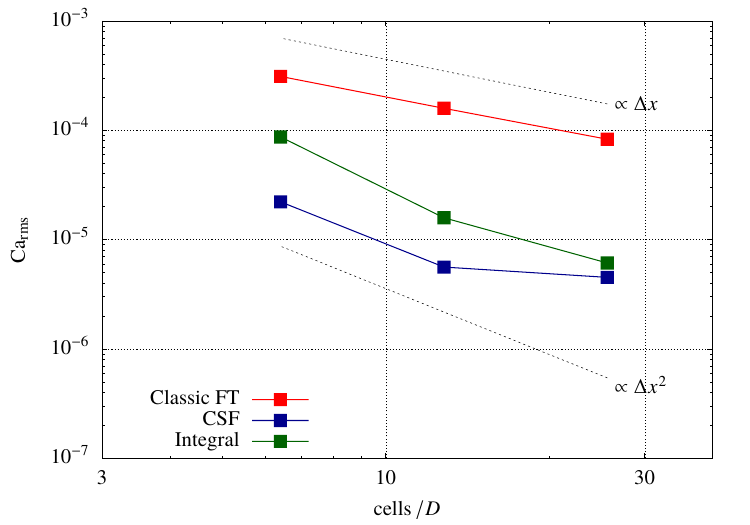}
    \caption{Analysis of the convergence rate of the root mean square capillary number for a moving droplet ($\mathrm{La = 12000}$, $\mathrm{We} = 0.4$), employing the classic FT, CSF and integral surface tension schemes. Results are shown in the droplet reference frame, and are averaged for $6.25 < tU_{\mathrm{ref},x}/D < 8$.}
    \label{fig:La12000_all_translating_convergence}
\end{figure}
The convergence rate of the root mean square capillary number for the three analysed surface tension schemes are compared in Figure \ref{fig:La12000_all_translating_convergence}, which shows that the classic approach converges with first--order accuracy, whereas the integral scheme is close to second--order accuracy. The CSF method exhibits second--order accuracy between the coarsest and intermediate resolutions, but it is not capable to further decrease spurious velocities at the highest resolution. This behaviour is attributed to the sensitivity of the CSF method to errors in the computation of the curvature. Remeshing operations and errors in the advection of the interface might cause local roughness on the front, which affects the PPIC reconstruction of the interface. This can induce noise in the curvature field and, in turn, local velocity fluctuations. Previous results based on dedicated local roughness smoothing techniques \citep{Gorges2025b} have shown a significant improvement in reducing numerical artefacts. However, it is reminded here that such techniques require configuration--dependent settings and are not used in the present work, where more general routines are preferred. On the other hand, the integral method, which still relies on quadratic fitting, exhibits consistent and close to second--order accuracy for all the tested resolutions. In this case, only the pressure correction factor $\beta$ depends on the curvature of the quadratic surface, and the eventual noise introduced by numerical errors does not compromise the global convergence rate.

\subsection{Oscillations of a slightly perturbed droplet}
In this test case, a droplet is initially perturbed along the $x$--direction and turned into a slightly ellipsoidal shape. As the deformed droplet is released in a quiescent flow, an oscillating dynamics takes place until viscous dissipation re--establishes an equilibrium solution and the droplet recovers a spherical shape. In order to capture the correct shape oscillations, an accurate estimation of the interfacial curvature is essential. This test cases is another widely adopted benchmark for surface tension schemes \citep{Gorges2025b, Abu-Al-Saud2018, Liu2023, Aalilija2020, deSousa2004}.

The analytical solution for a droplet with unperturbed radius $R_0$ under small perturbations was first provided by \citet{Lamb1932}:
\begin{equation}
    R(\theta, t) = R_0 + a_n P_n (\cos{\theta})\sin{\omega_n t} ,\quad\ \theta \in [0, 2\pi]
    \label{eq:lamb}
\end{equation}
where the radius of the droplet $(R)$ is formulated in a polar coordinate system $(\theta)$ centred on the droplet, whereas $P_n$ is the $n$--th Legendre polynomial, and $n$ refers to the $n$--th oscillation mode. The oscillation frequency $\omega_n$ reads:
\begin{equation}
    \omega_n = \sqrt{\frac{n (n+1) (n-1) (n+2) \sigma}{\left[ (n+1)\rho_d + n \rho_c \right] R_0^3}}
    \label{eq:lamb_omegan}
\end{equation}
The oscillation amplitude decays exponentially (see \citet{Liu2023} and the references therein):
\begin{equation}
    a_n(t) = a_0 \exp{(-\gamma t)} ,\quad\ \gamma = \frac{(n-1)(2n+1)\mu_d}{R_0^2 \rho_d}
    \label{eq:lamb_an}
\end{equation}

A droplet with unperturbed radius $R_0 = 1$ is placed at the centre of a cubic domain with size $4 R_0 \times 4R_0 \times 4R_0$, where the flow is initially at rest, and no gravitational acceleration is applied. The boundary conditions are set to symmetric walls, and the interface is deformed at $t=0$ according to Eq. \ref{eq:lamb}, with $n = 2$ and $a_0 = 0.025$. The $\theta = 0$ direction corresponds to the $x$--axis in our reference system. The problem is treated in a non--dimensional form, with density and viscosity ratios:
\begin{equation}
    \rho_\mathrm{r} = \frac{\rho_\mathrm{d}}{\rho_\mathrm{c}} = 100, \quad\ \mu_\mathrm{r} = \frac{\mu_\mathrm{d}}{\mu_\mathrm{c}} = 100,
\end{equation}
and two different Ohnesorge numbers:
\begin{equation}
    \mathrm{Oh} = \frac{\mu_d}{\sqrt{\rho_d \sigma R_0}} = 0.05, 0.005
\end{equation}
Results are compared against the analytical solution (Eq. \ref{eq:lamb}--\ref{eq:lamb_an}) for $R(\theta = 0, t)$. Measuring the radius as the distance between the markers coordinates and the droplet centre can result in small oscillations due to remeshing operations that locally alter the position of the front markers \citep{Gorges2025b}. In order to avoid such unphysical fluctuations, the time--evolving radius $R(\theta = 0, t)$ is computed as:
\begin{equation}
    R(\theta = 0, t) = \sqrt{\frac{5}{2 M} (\mathbf{I}_{yy} + \mathbf{I}_{zz} - \mathbf{I}_{xx})}
    \label{eq:ellipsoid}
\end{equation}
where $M$ is the mass of the droplet, and $\mathbf{I}$ is the inertia tensor. Eq. \ref{eq:ellipsoid} approximates the shape of the droplet with an ellipsoid, which is found to be accurate for the small perturbations considered in this work.

\begin{figure}[!htbp]
    \centering
    \includegraphics[]{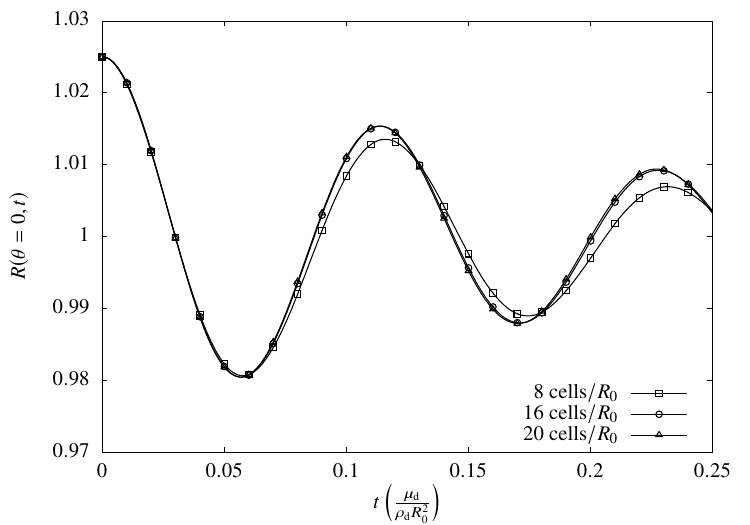}
    \caption{Mesh convergence analysis of the radius along the $x$--direction of a slightly perturbed oscillating droplet $(\mathrm{Oh} = 0.05)$. The simulations are run with the integral surface tension scheme.}
    \label{fig:oscillating_mesh_sensitivity}
\end{figure}
A mesh sensitivity study is first performed for the case with $\mathrm{Oh = 0.05}$, using the integral surface tension scheme. Figure \ref{fig:oscillating_mesh_sensitivity} shows the radius profiles for three different mesh resolution, namely 8, 16 and 20 cells per unperturbed radius $R_0$. The droplet radius, which initially is $R(\theta=0, t = 0) = 1.025$, decreases as a shrinkage mechanism is observed in the first half--period, until a local minimum is reached. An expansion mechanism follows, as the droplet releases the energy accumulated during the first part of the cycle. A periodic motion with damped oscillations characterises the droplet dynamics, consistent with the analytical solution reported above. The solver converges for a mesh resolution with 16 cells/$R_0$, as minimal differences are observed with the case at the highest resolution.

\begin{figure}[!htbp]
    \centering
    \includegraphics[]{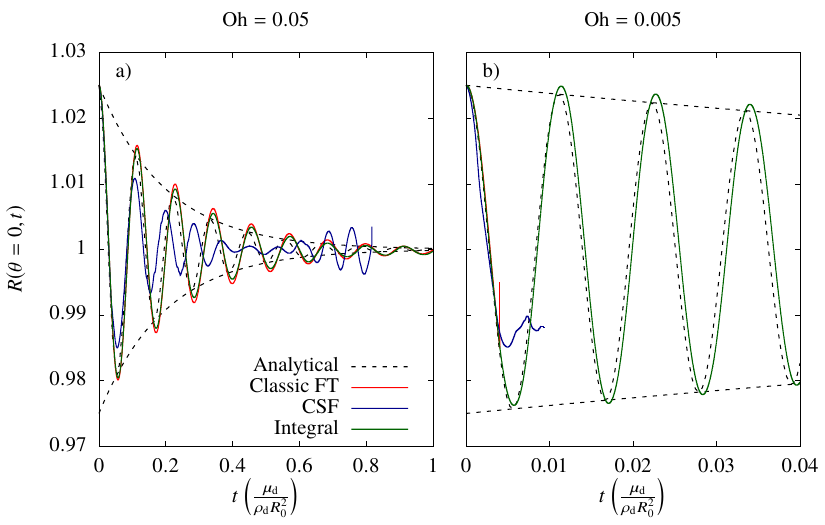}
    \caption{Time--evolving radius in the $x$--direction of slightly perturbed oscillating droplets with $\mathrm{Oh} = 0.05$ (a) and $\mathrm{Oh} = 0.005$ (b). The dotted lines represent the analytical solution for the radius (Eq. \ref{eq:lamb}) and the exponential decay of the amplitude (Eq. \ref{eq:lamb_an}).}
    \label{fig:oscillating}
\end{figure}
Classic FT, CSF and integral surface tension schemes are compared for two Ohnesorge numbers ($\mathrm{Oh} = 0.05, 0.005$), and mesh resolution of 16 cells/$R_0$, in Figure \ref{fig:oscillating}a and \ref{fig:oscillating}b, respectively. Classic FT and integral schemes perform similarly for $\mathrm{Oh} = 0.05$, where both approaches are in good agreement with the analytical solution, in terms of oscillation amplitude and frequency (although local maxima and minima with the classic FT scheme deviate slightly more from the reference solution). The simulations are run until the droplet recovers the equilibrium solution and the radius relaxes to $R(\theta=0, t) \approx R_0$. The CSF case, on the other hand, fails to capture the right oscillating dynamics and results in over--damped oscillations. It approaches the equilibrium solution for $0.4 < t(\mu_d/\rho_dR_0^2) <0.6$, before a new oscillating pattern takes place; the simulation diverges at $t(\mu_d/\rho_dR_0^2) \approx 0.8$. Such a behaviour is due to the effect of spurious velocities that generate local roughness on the front, inducing noise in the curvature field that, in turn, perturbs the solution via the surface tension source term. This result is consistent with our previous investigations \citep{Gorges2025b}, where this test case with the CSF method was found stable only when dedicated local roughness smoothing techniques are applied. It is reminded here that, in this work, only general--scope and globally--smoothing remeshing techniques, such as TSUR--3D and tangential relaxation, are used. 

The results of the second case $(\mathrm{Oh} = 0.005)$ are shown in Figure \ref{fig:oscillating}b. This test case is more demanding than the previous one, since it is characterised by a smaller dumping ratio, due to lower viscous dissipation rates. From a numerical perspective, this translates into increased spurious velocities. For comparison against the spherical droplet test cases with $\mathrm{La = 12000}$ (sections \ref{sec:Spurious velocities around a static spherical droplet}--\ref{sec:Spurious velocities around a translating spherical droplet}), the Laplace numbers corresponding to $\mathrm{Oh} = 0.05, 0.005$ are $\mathrm{La} = 800, 80000$, respectively (computed with $\rho_d, \mu_d$ in Eq.\ref{eq:laplace_number}). The results reported in Figure \ref{fig:oscillating}b show that only the integral method is able to converge to the analytical solution, with accurate predictions of both the amplitude and the frequency of the oscillations. On the other hand, spurious velocities are predominant with both classic FT and CSF schemes, and make the computations diverge almost immediately. This benchmark highlights the importance of testing numerical frameworks for surface tension driven flows under conditions where interfaces are not spherical and a static equilibrium between pressure gradient and surface tension is not reached.

\subsection{Migration of a droplet in a Marangoni flow}
The previous validations cases dealt with two--phase flow configurations where the surface tension coefficient was kept constant. In this section, we propose a test case in which $\sigma$ is not uniform, but varies linearly as a function of temperature:
\begin{equation}
    \sigma(T) = \sigma_0 + \sigma_T (T - T_0)
    \label{eq:sigma_marangoni_T}
\end{equation}
where $T_0$ is the reference temperature and $\sigma_0$ is the surface tension value for $T=T_0$, whereas $\sigma_T$ is a constant coefficient. For non--uniform surface tension coefficients (i.e., $\nabla_\Sigma \sigma \neq 0$), the balance of momentum at the interface (along a tangential direction $\mathbf{t}_\Sigma$) reads \citep{Tryggvason2011a}:
\begin{equation}
    -\left[ 2 \mu \mathbf{t}_\Sigma \cdot \mathbf{D} \cdot \mathbf{n}_\Sigma \right]_\Sigma = \mathbf{t}_\Sigma \cdot \nabla_\Sigma \sigma
\end{equation}
where the term on the LHS refers to the jump of tangential shear stresses across the interface, and $\mathbf{D}$ is the deformation tensor (Eq. \ref{eq:momentum}). Therefore, a variable surface tension coefficient along the interface can be balanced only by viscous shear stresses, which in turn drive the motion of the fluid. Such a flow configuration is known as Marangoni flow. Due to the dependency of $\sigma$ on the temperature field, this configuration is also referred to as thermocapillary flow, which forces the droplet to move from the cold region to the hot one.

The test cases proposed in this section are adapted from \citep{Abu-Al-Saud2018, Saini2025a}. A droplet with radius $R = 1$ is placed in a cubic domain with size $32R \times 32R \times 32R$, where no gravitational acceleration is applied. The temperature field is kept constant during the simulation (i.e., no transport equation for $T$ is solved), and varies linearly in the $x$--direction as:
\begin{equation}
    T(x) = T_0 + \frac{dT}{dx}x
\end{equation}
where the temperature gradient is assumed constant. The surface tension coefficient (Eq. \ref{eq:sigma_marangoni_T}) can be rewritten as:
\begin{equation}
    \sigma(x) = \sigma_0 + \sigma_T \frac{dT}{dx}x
\end{equation}
The analytical solution for the steady--state droplet velocity was provided by \citet{Young1959} and reads:
\begin{equation}
    U_\mathrm{analytical} = - \frac{2}{(2 + 3 \mu_\mathrm{d}/\mu_\mathrm{c})(2 + c_\mathrm{d}/c_\mathrm{c})}\frac{\sigma_T R (dT/dx)}{\mu_\mathrm{c}}
\end{equation}
where $c_\mathrm{d}, c_\mathrm{c}$ are the thermal conductivity of the disperse and continuous phases, respectively, and their ratio is set to one. Under these assumptions, the non--dimensional problem is defined by the density and viscosity ratios:
\begin{equation}
    \rho_\mathrm{r} = \frac{\rho_\mathrm{d}}{\rho_\mathrm{c}} = 1, \quad\ \mu_\mathrm{r} = \frac{\mu_\mathrm{d}}{\mu_\mathrm{c}} = 1, 0.5 \quad\ \text{(two different viscosity ratios are tested)},
\end{equation}
the Reynolds number:
\begin{equation}
    \mathrm{Re} = \frac{\rho_c U_\mathrm{ref} R}{\mu_c} = 0.066,
\end{equation}
and the capillary number:
\begin{equation}
    \mathrm{Ca} = \frac{\mu_c U_\mathrm{ref}}{\sigma_0} = 0.066,
\end{equation}
where the reference velocity is defined as:
\begin{equation}
    U_\mathrm{ref} = \frac{\sigma_T R}{\mu_c}\frac{dT}{dx}
\end{equation}

\begin{figure}[!htbp]
    \centering
    \includegraphics[]{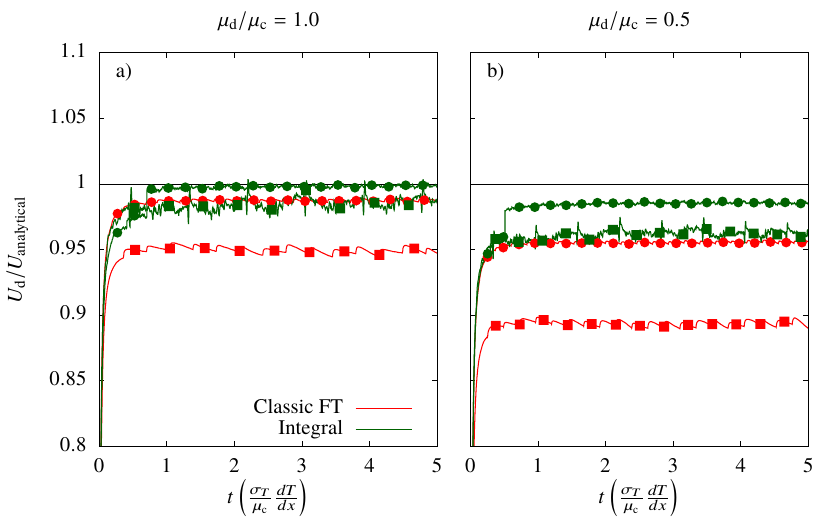}
    \caption{Translating velocity of a droplet in a thermocapillary flow with $\mathrm{Re} = 0.066$, $\mathrm{Ca} = 0.066$ and viscosity ratios $\mu_d/\mu_c = 1$ (a) and $\mu_d/\mu_c = 0.5$ (b). Grid resolutions: 9 cells$/R$ (---$\blacksquare$---) and 18 cells$/R$ (---$\bullet$---).}
    \label{fig:thermal_marangoni}
\end{figure}

\begin{figure}[!htbp]
    \centering
    \includegraphics[]{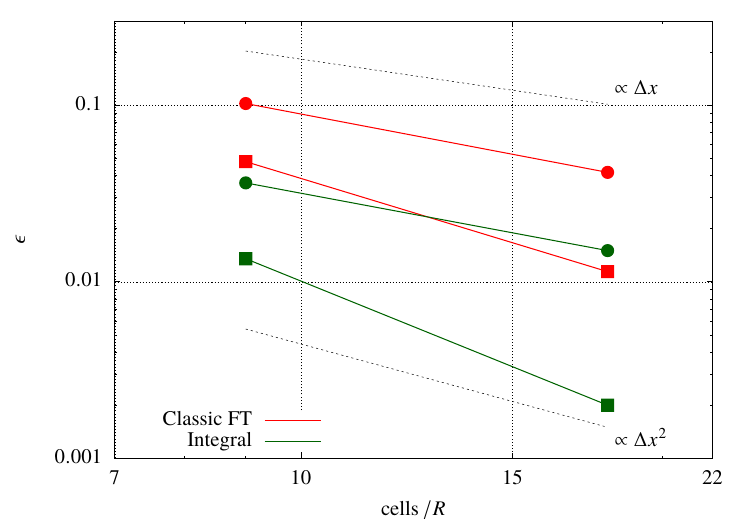}
    \caption{Analysis of the convergence rate of the terminal translating velocity of a droplet in a thermocapillary flow with $\mathrm{Re} = 0.066$, $\mathrm{Ca} = 0.066$, and two different viscosity ratios: $\mu_\mathrm{r} = 1$ (---$\blacksquare$---) and $\mu_\mathrm{r} = 0.5$ (---$\bullet$---).}
    \label{fig:thermal_marangoni_convergence}
\end{figure}

The results for the translating velocity of the droplet $U_\mathrm{d}$ are shown in Figure \ref{fig:thermal_marangoni}a and \ref{fig:thermal_marangoni}b for the viscosity ratios $\mu_\mathrm{r} = 1$ and $\mu_\mathrm{r} = 0.5$, respectively. 
For this specific test cases, only the classic FT and integral surface tension schemes are tested (with resolutions corresponding to 9 and 18 cells per droplet radius), since these offer a seamless treatment for cases with variable $\sigma$. For both viscosity ratios and mesh resolutions, the integral method outperforms the classic FT approach and provides very accurate predictions of the terminal velocity, with larger discrepancies for the cases with non--uniform viscosities (consistent with the results reported in \citep{Abu-Al-Saud2018}). A quantitative comparison of the relative error:
\begin{equation}
    \epsilon = \frac{|U_\mathrm{analytical} - U_\mathrm{d}|}{U_\mathrm{analytical}}
\end{equation}
is reported in Table \ref{tab:marangoni_error},
\begin{table}[htbp]
\centering
\begin{tabular}{r cccc}
\toprule
& \multicolumn{2}{c}{Classic FT} & \multicolumn{2}{c}{Integral} \\
\cmidrule(lr){2-3} \cmidrule(lr){4-5}
$R/\Delta$ & $\mu_{\mathrm{d}}/\mu_{\mathrm{c}} = 1$ & $\mu_{\mathrm{d}}/\mu_{\mathrm{c}} = 0.5$ & $\mu_{\mathrm{d}}/\mu_{\mathrm{c}} = 1$ & $\mu_{\mathrm{d}}/\mu_{\mathrm{c}} = 0.5$ \\
\midrule
9  & 0.0479 & 0.1021 & 0.0135 & 0.0363 \\
18 & 0.0114 & 0.0416 & 0.0020 & 0.0150 \\
\bottomrule
\end{tabular}
\caption{Relative errors for the translating velocity of a droplet in a thermocapillary flow with $\mathrm{Re} = 0.066$ and $\mathrm{Ca} = 0.066$, for the classic FT and integral surface tension schemes with two different mesh resolutions.}
\label{tab:marangoni_error}
\end{table}
and the corresponding convergence plot is shown in Figure \ref{fig:thermal_marangoni_convergence}.
For the most challenging case with $\mu_\mathrm{r} = 0.5$, the integral surface tension scheme achieves a convergence rate between first and second order, whereas the classic FT approach converges with less than first order accuracy.

\begin{figure}[!htbp]
    \centering
    \includegraphics[]{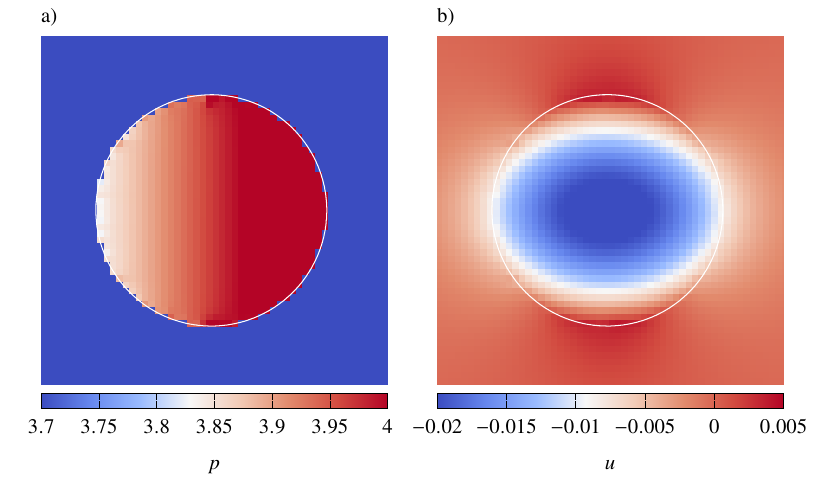}
    \caption{Pressure (a) and $x$--component $u$ of the velocity field (b) of a droplet in a thermocapillary flow with $\mathrm{Re} = 0.066$, $\mathrm{Ca} = 0.066$, and $\mu_\mathrm{r} = 1$. The mesh resolution is 18 cells$/R$, and the surface tension scheme is the integral one. The temperature gradient is set to $dT/dx = -1$.}
    \label{fig:thermal_marangoni_contour}
\end{figure}

A qualitative representation of the flow field is shown in Figure \ref{fig:thermal_marangoni_contour}a and \ref{fig:thermal_marangoni_contour}b for the pressure and the $x$--velocity component, respectively, for the case with unit viscosity ratio, largest mesh resolution and integral surface tension. 
The temperature gradient is negative and, as a result of the variable surface tension field, it generates a negative pressure gradient in the $x$--direction (Figure \ref{fig:thermal_marangoni_contour}a) that pushes the droplet to migrate from the cold region towards the hot one (Figure \ref{fig:thermal_marangoni_contour}b).

\subsection{Rising bubbles}
In the test cases discussed above, the system is entirely driven by surface tension, as no other source terms are applied. In this section, we apply the proposed numerical framework to bubbles rising in a quiescent liquid, where the evolution of the interface results from the combination of buoyancy and surface tension effects. 

A spherical bubble with diameter $D=1$ is placed in a liquid domain with dimensions $10D \times 10D \times 15D$, where the gravitational acceleration $(\mathbf{g} = [0, 0, -g])$ acts along the $z$--direction only. No--slip boundary conditions are applied everywhere except at the top boundary, where a pressure--outlet condition is set. An adaptive mesh refinement (AMR) approach (based on the location of the interface and flow vorticity) is employed to reduce the computational cost of the simulations, whilst ensuring that the relevant flow scales around the bubble and in the wake region are properly resolved. Surface tension is assumed uniform along the interface, and the non-dimensional form of the problem is completely defined by the density and viscosity ratios:
\begin{equation}
    \rho_\mathrm{r} = \frac{\rho_\mathrm{d}}{\rho_\mathrm{c}}, \quad\ \mu_\mathrm{r} = \frac{\mu_\mathrm{d}}{\mu_\mathrm{c}},
\end{equation}
the Bond number:
\begin{equation}
    \mathrm{Bo} = \frac{g D^2 \rho_\mathrm{c}}{\sigma},
\end{equation}
and the Morton number:
\begin{equation}
    \mathrm{Mo} = \frac{g \mu_\mathrm{c}^4}{\rho_\mathrm{c} \sigma^3}
\end{equation}

In this section, we reproduce three test cases from the experimental work of \citet{Bhaga1981}, whose non--dimensional numbers are reported in Table \ref{tab:rising}.
\begin{table}[htbp]
\centering
\begin{tabular}{l cccc}
\toprule
Case &$\rho_\mathrm{r}$ & $\mu_\mathrm{r}$ & $\mathrm{Bo}$ & $\mathrm{Mo}$ \\
\midrule
Case 1 &$9.0\times10^{-4}$ & $6.5\times10^{-4}$ & 116 & 848 \\
Case 2 &$9.0\times10^{-4}$ & $1.4\times10^{-3}$ & 116 & 41.1 \\
Case 3 &$9.0\times10^{-4}$ & $3.3\times10^{-3}$ & 116 & 1.31 \\
\bottomrule
\end{tabular}
\caption{Non--dimensional numbers for three different cases of rising bubbles.}
\label{tab:rising}
\end{table}
The results are compared against the experimental measurements \citep{Bhaga1981} in terms of terminal rising velocities $(U_\mathrm{rising})$ via the bubble Reynolds number:
\begin{equation}
    \mathrm{Re} = \frac{\rho_\mathrm{c} U_\mathrm{rising} D}{\mu_\mathrm{c}}
\end{equation}

\begin{figure}[!htbp]
    \centering
    \includegraphics[]{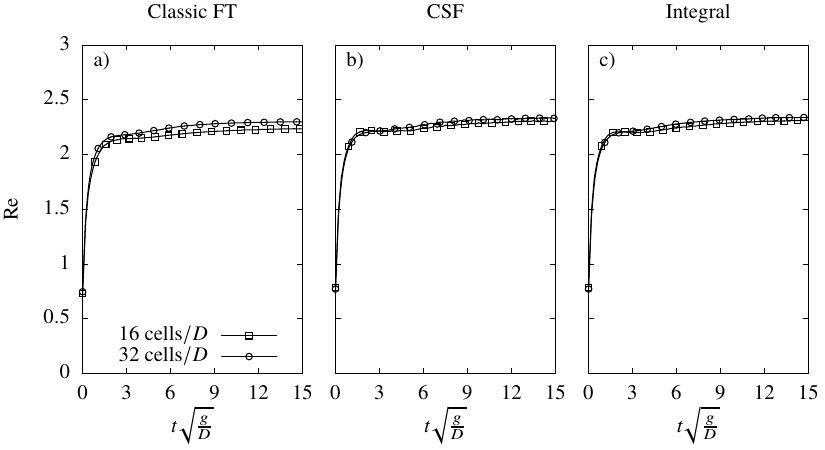}
    \caption{Mesh sensitivity analysis for case 1 of rising bubbles (see Table \ref{tab:rising}) using the classic FT (a), CSF (b) and integral (c) surface tension schemes.}
    \label{fig:rising_mesh_sensitivity}
\end{figure}

A mesh sensitivity study is first performed for Case 1 $(\mathrm{Mo} = 848)$ and shown in Figure \ref{fig:rising_mesh_sensitivity}.
Two mesh resolutions, corresponding to 16 and 32 cells per bubble diameter, are compared for all the three surface tension schemes. In all the cases, small differences for steady--state values are observed between the two resolutions ($\approx 2.5\%$ for the classic scheme, and $\approx 1\%$ for the CSF and integral approaches). Given the little increase in computational cost achieved with the use of AMR, the mesh resolution with 32 cells per bubble diameter is selected for the following analysis.

\begin{figure}[!htbp]
    \centering
    \includegraphics[]{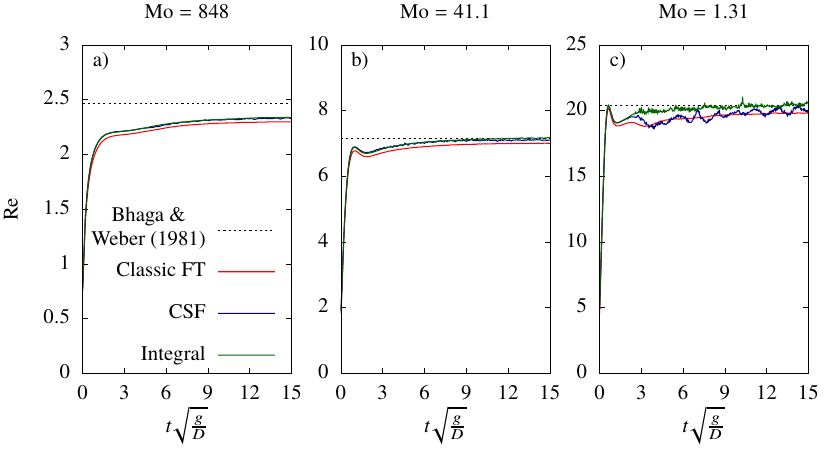}
    \caption{Bubbles Reynolds numbers for case 1 (a), case 2 (b) and case 3 (c) (see Table \ref{tab:rising}) of rising bubbles, employing the classic FT, CSF and integral surface tension schemes. The results are compared against the steady--state measurements from the experiments of \citet{Bhaga1981}.}
    \label{fig:rising_all}
\end{figure}

\begin{figure}[!htbp]
    \centering
    \includegraphics[]{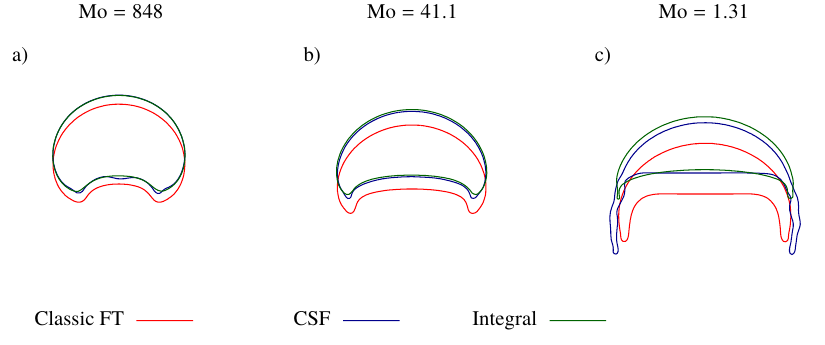}
    \caption{Steady--state rising bubble shapes (2D slices on the $xy$--plane) computed at $t \approx 15 \sqrt{D/g}$ for the three cases reported in Table \ref{tab:rising}, employing the classic FT, CSF and integral surface tension schemes.}
    \label{fig:rising_shapes}
\end{figure}

The terminal bubble Reynolds numbers are compared in Figure \ref{fig:rising_all} against the corresponding experimental values for all the cases reported in Table \ref{tab:rising}. 
A quantitative analysis based on the relative error between numerical simulations and experiments:
\begin{equation}
    \epsilon = \frac{|\mathrm{Re_{exp}} - \mathrm{Re}|}{\mathrm{Re_{exp}}}
    \label{eq:err_Re}
\end{equation}
is reported in Table \ref{tab:rising_error}.
\begin{table}[htbp]
\centering
\begin{tabular}{l ccc}
\toprule
Case &Classic FT &CSF &Integral \\
\midrule
Case 1 &$7.04\%$ & $5.60\%$ & $5.41\%$\\
Case 2 &$2.14\%$ & $0.73\%$ & $0.08\%$\\
Case 3 &$2.87\%$ & $1.85\%$ & $0.16\%$\\
\bottomrule
\end{tabular}
\caption{Relative error between experiments \citep{Bhaga1981} and numerical simulations for the bubble Reynolds numbers (Eq. \ref{eq:err_Re}), using the classic FT, CSF and integral surface tension schemes. The numerical values are averaged in the time interval $12 < t\sqrt{g/D} < 15$.}
\label{tab:rising_error}
\end{table}
Overall, the integral surface tension scheme achieves the lowest deviations from the experiments, followed by the CSF method, whereas the classic framework results in larger differences. Case 1 has the larger deviations from the experimentally--measured Reynolds number, but the reported values for the CSF and integral methods are close to the uncertainty in the experimental measurements, which is reported to be approximately $\approx 5\%$. Notably, the integral method provides very accurate results for case 3, where the bubble undergoes strong deformations (see Figure \ref{fig:rising_shapes}c). Given the reported measurement error, we conclude that the CSF and integral results are consistent with all the experiments analysed in this section, although the integral scheme results in better quantitative predictions.

A qualitative comparison of the bubble shapes at $t \approx 15 \sqrt{D/g}$ is reported in Figure \ref{fig:rising_shapes}.
No appreciable differences (except for the position, due to underestimated rising velocities with the classic FT scheme) can be observed for cases 1 (Figure \ref{fig:rising_shapes}a) and 2 (Figure \ref{fig:rising_shapes}b). Stronger deformations and differences between the surface tension treatments are observed for case 3, which is characterised by the lowest Morton number, and results in the formation of a trailing skirt behind a spherical cap for the classic FT and CSF methods. The thickness of the skirt is the largest with the classic scheme. Consistently, limited capabilities in terms of capturing thin interfacial structures with the classic FT framework are also observed in \citep{Gorges2025b}. However, substantial differences are evident when comparing these results against the integral method, which returns a bubble shape characterised by a tiny elongation of the skirt structure. Similarly, the experiments of \citet{Bhaga1981}, as well as other numerical studies that reproduce the same case \citep{Liu2023, Anjos2014}, report steady-state bubble shapes characterised by very short elongations of the skirt. The proposed integral method, therefore, results in very good agreement with the reference data (both quantitatively and qualitatively), and clearly outperforms the classic FT and CSF schemes for rising bubbles at low Morton numbers.

\section{Conclusions}
\label{sec:Conclusions}
In this work, we have developed the first three--dimensional integral surface tension scheme built on the original 2D work of \citet{Popinet1999} and later developments \citep{Abu-Al-Saud2018, Saini2025a}. The integral scheme has the advantage of conserving momentum both locally and globally, and offers a natural numerical framework to treat two--phase systems with variable surface tension coefficients, such as Marangoni flows. Surface tension is computed following a sharp approach, which consists of applying the integral definition of surface tension at the intersections between the interface and the faces of each computational cell. An explicit pressure correction is applied to take into account the pressure jump across the interface, following the Laplace pressure equilibrium condition. This approach results in a non--null source term only for those cells that are actually cut by the interface, whereas every other cell has null surface tension contributions, returning a truly sharp implementation of the interfacial force. The proposed scheme is implemented in a state--of--the--art sharp front--tracking framework \citep{Gorges2025b} and makes use of a PPIC reconstruction method to compute the required intersections between the interface and the computational cells. The explicit representation of the interface in the FT method facilitates the use of the PPIC reconstruction, providing very accurate approximations of the interface and robust computations of the intersections, even in very complex three--dimensional configurations. 

The proposed numerical framework is compared against well--known surface tension schemes adopted in the literature, such as the CSF method and the classic (smoothing--based) approach commonly employed in FT simulations. The effectiveness of the methodology is tested across a variety of benchmarks, including surface tension dominated systems (e.g., static and translating spherical droplets, oscillating droplets and Marangoni flows) and rising bubbles. The results are compared against both analytical solutions and experimental data, showing overall superior performance of the integral scheme, with the largest gains achieved for large deformations in rising bubbles or in surface tension dominated flows, such as oscillating droplets at very low Ohnesorge numbers or two--phase systems involving Marangoni flows. Droplets that oscillate at low $\mathrm{Oh}$ numbers are more sensitive to spurious velocities, since capillary effects are dominant over viscous dissipation. The integral surface tension scheme is the only one that remains numerically stable for $\mathrm{Oh = 0.005}$, without relying on case--dependent remeshing techniques to mitigate numerical noise, and is able to capture the corresponding analytical solution describing the damping of the oscillations. The integral method is particularly suited for two--phase flows with variable surface tension coefficients. The simulations of the thermocapillary motion of a droplet show a reduction in the error of the predicted terminal velocity up to an order of five compared to the classic (smoothing--based) approach for FT. When modelling large deformations of the interface, the integral scheme outperforms both CFS and classic FT. For rising bubbles at low Morton numbers, the integral method achieves a very accurate prediction of the terminal velocity, with a deviation of only 0.16\% from the experiments, whereas CSF and classic FT produce 1.85\% and 2.87\% errors, respectively. The predicted shape of steady--state rising bubbles undergoing large deformations is in close agreement with the experiments when the proposed surface tension scheme is employed, whereas CSF and classic FT return significantly different shapes.

\section*{Acknowledgements}
\noindent This project has received funding from the Deutsche Forschungsgemeinschaft (DFG, German Research Foundation), grant numbers 420239128 and 458610925. The authors thank Christian Gorges for fruitful discussions.

\section*{Data availability}
\noindent
The data that support the findings of this study are openly available and can be downloaded from the repositories \href{https://doi.org/10.5281/zenodo.21300906}{https://doi.org/10.5281/zenodo.21300906} and
\href{https://doi.org/10.5281/zenodo.21301616}{https://doi.org/10.5281/zenodo.21301616}.

\bibliographystyle{model1-num-names}

\end{document}